\documentclass[12pt]{article}

\usepackage{array}
\usepackage{amssymb}
\usepackage{amsmath}
\usepackage{ntheorem}
\usepackage{epsf}
\usepackage{color}
\usepackage{xcolor}
\usepackage{graphicx}
\usepackage{epstopdf}
\usepackage{subfigure}
\usepackage{arydshln}
\usepackage{makecell}
\usepackage{mathrsfs}
\usepackage{multirow}
\usepackage{longtable}
\usepackage{threeparttable}
\usepackage{url}
\usepackage{makecell}
\usepackage{booktabs}
\usepackage{listings}
\usepackage{float}
\usepackage{diagbox}
\usepackage[bookmarks=false]{hyperref}
\hypersetup{hidelinks,
	colorlinks=true,
	allcolors=blue,
	pdfstartview=Fit,
	breaklinks=true}

\usepackage[ruled, linesnumbered]{algorithm2e}
\def\bfm#1{\mbox{\boldmath$#1$}}
\numberwithin{equation}{section}

\usepackage[style=authoryear, giveninits=true, uniquename=init, backend=biber, maxbibnames=99, minbibnames=3, maxcitenames=2]{biblatex}

\DeclareFieldFormat[article]{volume}{\textbf{#1}}

\DeclareFieldFormat[article,inbook,incollection,inproceedings]{title}{#1}

\renewbibmacro{in:}{%
  \ifboolexpr{test {\ifentrytype{incollection}} or test {\ifentrytype{inproceedings}}}
    {\printtext{In\space}}
    {}%
}
\DeclareFieldFormat[article]{number}{\mkbibparens{#1}}

\renewbibmacro*{volume+number+eid}{%
  \printfield{volume}%
  \printfield{number}% 这里直接跟着期号，不加任何点或空格
  \setunit{\addcomma\space}%
  \printfield{eid}}
\DeclareFieldFormat[article]{pages}{#1}
\DeclareNameAlias{sortname}{family-given}

\renewbibmacro*{name:andothers}{\ifboolexpr{test {\ifnumequal{\value{listcount}}{\value{liststop}}} and test \ifmorenames}{\ifnumgreater{\value{liststop}}{1}{\finalandcomma}{}\printdelim{andothersdelim}\textit{et al.}}{}}
\DeclareFieldFormat{doi}{https://doi.org/#1}
\DeclareFieldFormat{doi}{\url{https://doi.org/#1}}

\renewbibmacro*{addendum+pubstate}{%
  \printfield{addendum}%
  \iffieldundef{pubstate}
    {}
    {\setunit{\addcomma\space}% 强制将前面的标点改为逗号
     \printfield{pubstate}}%
}

\DeclareFieldFormat[book]{volume}{#1}
\DeclareFieldFormat[book]{edition}{#1}

\renewbibmacro*{title}{%
  \ifboolexpr{test {\iffieldundef{title}} and test {\iffieldundef{subtitle}}}
    {}
    {\printtext[title]{%
       \printfield[titlecase]{title}%
       \setunit{\subtitlepunct}%
       \printfield[titlecase]{subtitle}}%
     \ifentrytype{book}
       {\setunit{\addspace}% 【如果是书籍】加空格，准备拼接括号
        \ifboolexpr{test {\iffieldundef{volume}} and test {\iffieldundef{edition}}}
          {} 
          {\printtext[parens]{% 加上小括号
             \iffieldundef{volume}{}{Vol.\addspace\printfield{volume}}%
             \ifboolexpr{not test {\iffieldundef{volume}} and not test {\iffieldundef{edition}}}
               {\addcomma\space}{}%
             \iffieldundef{edition}{}{\printfield{edition}\addspace Edition}%
           }}%
        \clearfield{volume}% 仅清除书籍的卷号，绝不碰期刊的卷号！
        \clearfield{edition}% 
       }
       {}% 【如果不是书籍】（如期刊Article），什么都不做，安全退出
     \newunit}%
}

\renewbibmacro*{publisher+location+date}{%
  \printlist{publisher}% 先打印出版社
  \setunit*{\addcomma\space}% 插入“逗号+空格”
  \printlist{location}% 再打印出版地
  \newunit}

\DeclareFieldFormat[incollection]{pages}{#1}

\renewbibmacro*{booktitle}{%
  \ifboolexpr{test {\iffieldundef{booktitle}} and test {\iffieldundef{booksubtitle}}}
    {}
    {\printtext[booktitle]{%
       \printfield[titlecase]{booktitle}%
       \setunit{\subtitlepunct}%
       \printfield[titlecase]{booksubtitle}}%
     \iffieldundef{pages}
       {}
       {\setunit{\addcomma\space}% 插入逗号和空格
        \printfield{pages}% 打印页码
        \clearfield{pages}}% 关键：打印完立刻销毁，防止结尾重复打印
    }%
}

\DeclareFieldFormat[unpublished,inbook,incollection,inproceedings, misc]{title}{#1}
\begin{document}

% 0. Example et al

%\renewcommand{\theequation}{\thesection.\arabic {equation}}
\theoremstyle{plain}
\newtheorem{corollary}{Corollary}[section]
\newtheorem{definition}{Definition}[section]
\newtheorem{example}{Example}[section]
\newtheorem{lemma}{Lemma}%[section]
\newtheorem{property}{Property}[section]
\newtheorem{remark}{Remark}[section]
\newtheorem{theorem}{Theorem}%[section]
\newtheorem{Table}{Table}
\newtheorem{proof}{Proof}
%\renewcommand{\qedsymbol}{}

% 1. Empty-style

\newcommand{\bbB}{{\mathbb{B}}}
\newcommand{\bbC}{{\mathbb{C}}}
\newcommand{\bbD}{{\mathbb{D}}}
\newcommand{\bbE}{{\mathbb{E}}}
\newcommand{\bbI}{{\mathbb{I}}}
\newcommand{\bbJ}{{\mathbb{J}}}
\newcommand{\bbK}{{\mathbb{K}}}
\newcommand{\bbN}{{\mathbb{N}}}
\newcommand{\bbO}{{\mathbb{O}}}
\newcommand{\bbP}{{\mathbb{P}}}
\newcommand{\bbR}{{\mathbb{R}}}
\newcommand{\bbS}{{\mathbb{S}}}
\newcommand{\bbT}{{\mathbb{T}}}
\newcommand{\bbV}{{\mathbb{V}}}
\newcommand{\bbX}{{\mathbb{X}}}
\newcommand{\bbY}{{\mathbb{Y}}}

% 2. Calligraphic

\newcommand{\cA}{{\cal A}}         \newcommand{\dA}[1]{{\cal A}_{#1}}
\newcommand{\cB}{{\cal B}}         \newcommand{\dB}[1]{{\cal B}_{#1}}
\newcommand{\cI}{{\cal I}}
\newcommand{\cJ}{{\cal J}}
\newcommand{\cK}{{\cal K}}
\newcommand{\cN}{{\cal N}}
\newcommand{\cO}{{\cal O}}
\newcommand{\cP}{{\cal P}}
\newcommand{\cS}{{\cal S}}         \newcommand{\dS}[1]{{\cal S}_{#1}}
\newcommand{\cR}{{\cal R}}
\newcommand{\cT}{{\cal T}}
\newcommand{\cU}{{\cal U}}
\newcommand{\cV}{{\cal V}}
\newcommand{\cX}{{\cal X}}
\newcommand{\cY}{{\cal Y}}
\newcommand{\cZ}{{\cal Z}}

% 3. Function

\newcommand{\f}[1]{f_{_{\tiny\rm #1}}}  \newcommand{\hf}[1]{\hat{f}_{#1}}
\newcommand{\tf}[2]{f_{{#1}}^{{#2}}}

\newcommand{\h}[1]{h_{_{\tiny\rm #1}}}
\newcommand{\n}[1]{n_{_{\tiny\rm #1}}}
\newcommand{\q}[2]{q_{_{\tiny\rm #1\!, \, #2}}}
\newcommand{\del}[1]{\delta_{_{\tiny\rm #1}}}
\newcommand{\pii}[2]{\pi_{_{\tiny\rm #1\!, \, #2}}}

\newcommand{\y}[2]{y_{#1}^{\rm #2}}
\newcommand{\yb}[1]{{\bar{y}^{\rm #1}}}
\newcommand{\Y}[1]{Y_{\rm #1}}
\newcommand{\YY}[1]{Y_{\! \mbox{\scriptsize #1}}}
\newcommand{\Two}[2]{{#1}_{_{\textrm{#2}}}}

\newcommand{\no}[2] {   \| #1 \|_{_{#2}}   }
\newcommand{\nor}[3]{\|#1\|_{_{#2}}^{#3}}

\newcommand{\bismall}[2]{(\!\!\begin{array}{c} #1 \\ \vspace*{-0.6cm} \\ #2 \end{array} \!\! )}
\newcommand{\bi}[2]{\Big(\!\!\begin{array}{c} #1 \\ #2 \end{array} \!\!\Big)}
\newcommand{\twoc}[2]{\bigg(\!\!\begin{array}{c} #1 \\ #2  \end{array} \!\!\bigg)}
\newcommand{\twolr}[2]{\left(\!\!\begin{array}{c} #1 \\ #2  \end{array} \!\!\right)}

\newcommand{\twob}[2]{\lbrace {#1 \atop #2} \rbrace}
\newcommand{\twoB}[2]{\bigg\{\!\!\begin{array}{c} #1 \\ #2  \end{array} \!\!\bigg\}}

\newcommand{\fourc}[4]{\bigg(\!\!\begin{array}{cc} #1 & #2 \\ #3 & #4 \end{array} \!\!\bigg)}
\newcommand{\fourlr}[4]{\left(\!\!\begin{array}{cc} #1 & #2 \\ #3 & #4 \end{array} \!\!\right)}
\newcommand{\Cases}[4]{\left\{\!\!\begin{array}{ll} #1, & #2,  \\ [3mm]
                                                    #3, & #4  \end{array} \right. }

\newcommand{\threec}[3]{\left(\!\!\begin{array}{c} #1 \\ #2 \\ #3  \end{array} \!\!\right)}
\newcommand{\threetwo}[6]{\left(\!\!\begin{array}{cc}
#1 & #2 \\
#3 & #4 \\
#5 & #6
\end{array} \!\!\right)}

\newcommand{\threethree}[9]{\left(\!\!\begin{array}{ccc}
#1 & #2 & #3 \\
#4 & #5 & #6 \\
#7 & #8 & #9
\end{array} \!\!\right)}

\newcommand{\sr}[1]{\stackrel{#1}{=}}
\newcommand{\srto}[1]{\stackrel{\rm #1}{\to}}
\newcommand{\sle}[1]{\stackrel{#1}{\le}}
\newcommand{\sgr}[1]{\stackrel{#1}{\ge}}
\newcommand{\sge}[1]{\stackrel{#1}{\ge}}
\newcommand{\mc}[1]{\multicolumn{1}{c}{#1}}
\newcommand{\MC}[3]{\multicolumn{#1}{#2}{#3}}
\newcommand{\MR}[3]{\multirow{#1}{#2}{#3}}
\newcommand{\TC}[4]{\contentsline {#1}{\numberline {#2}#3}{#4}}
\newcommand{\LIST}[3]{\contentsline {section}{\numberline {\hspace*{-0.5cm}#1}#2}{#3}}

\newcommand{\Hpi}[1]{\hat{\pi}_{_{\tiny\rm #1}}}
\newcommand{\titem}[1]{\vspace*{-0.15cm}\item[{\rm #1} ]\hspace*{0.1cm}}
\newcommand{\ub}[1]{\underline{\bf #1}}
\newcommand{\RED}[1]{{\color{red} #1}}      \newcommand{\redbf}[1]{{\color{red} \bf #1}}
\newcommand{\GREEN}[1]{{\color{green} #1}}
\newcommand{\BLUE}[1]{{\color{blue} #1}}

\newcommand{\two}[2]{{#1}_{_{\tiny\rm #2}}}
\newcommand{\three}[3]{{#1}_{_{#2, \tiny\rm #3}}}

\newcommand{\threev}[3]{\left(\!\!\begin{array}{c} #1 \\   #2 \\   #3 \end{array} \!\!\right)}

\newcommand{\at}[1]{a^{(t)}_{#1}}
\newcommand{\bt}[1]{b^{(t)}_{#1}}
\newcommand{\ct}[1]{c^{(t)}_{#1}}       \newcommand{\Ct}[1]{C^{(t)}_{#1}}
\newcommand{\dt}[1]{d^{(t)}_{#1}}
\newcommand{\gt}[1]{g^{(t)}_{#1}}
\newcommand{\pt}[1]{p^{(t)}_{#1}}  \newcommand{\ibpt}[1]{\ibp^{(t)}_{#1}}
\newcommand{\rt}[1]{r^{(t)}_{#1}}
\newcommand{\st}[1]{s^{(t)}_{#1}}

% 4. bold 0 & 1

\newcommand{\0}{{\bf 0}\!\!\!{\bf 0}}
\newcommand{\1}{{\bf 1}\!\!\!{\bf 1}}        \def\b1{{\bf 1\!\!\!1}}  % Letter + number
\newcommand{\2}{{\bf 2}\!\!\!{\bf 2}}        \def\b2{{\bf 2\!\!\!2}}  % Letter + number

% 5. bold English Letter

% 5.1    lowercase
\newcommand{\ba}{{\bf a}}
\newcommand{\bb}{{\bf b}}
\newcommand{\bd}{{\bf d}}
\newcommand{\bn}{{\bf n}}
\newcommand{\bm}{{\bf m}}
\newcommand{\bp}{{\bf p}}
\newcommand{\bq}{{\bf q}}
\newcommand{\br}{{\bf r}}
\newcommand{\bs}{{\bf s}}
\newcommand{\bu}{{\bf u}}
\newcommand{\bv}{{\bf v}}
\newcommand{\bw}{{\bf w}}
\newcommand{\bx}{{\bf x}}
\newcommand{\by}{{\bf y}}
\newcommand{\bz}{{\bf z}}

% 5.2    uppercase
\newcommand{\bA}{{\bf A}}
\newcommand{\bB}{{\bf B}}
\newcommand{\bC}{{\bf C}}
\newcommand{\bD}{{\bf D}}
\newcommand{\bG}{{\bf G}}
\newcommand{\bH}{{\bf H}}
\newcommand{\bI}{{\bf I}}
\newcommand{\bJ}{{\bf J}}
\newcommand{\bL}{{\bf L}}
\newcommand{\bM}{{\bf M}}
\newcommand{\bO}{{\bf O}}
\newcommand{\bP}{{\bf P}}
\newcommand{\bQ}{{\bf Q}}
\newcommand{\bR}{{\bf R}}
\newcommand{\bS}{{\bf S}}
\newcommand{\bT}{{\bf T}}
\newcommand{\bU}{{\bf U}}
\newcommand{\bV}{{\bf V}}
\newcommand{\bW}{{\bf W}}
\newcommand{\bX}{{\bf X}}
\newcommand{\bY}{{\bf Y}}
\newcommand{\bZ}{{\bf Z}}

% 5.3 italic English Letter

\newcommand{\iba}{{\boldsymbol{a}}}    \newcommand{\ibA}{{\boldsymbol{A}}}
\newcommand{\ibb}{{\boldsymbol{b}}}    \newcommand{\ibB}{{\boldsymbol{B}}}
                                       \newcommand{\ibC}{{\boldsymbol{C}}}
                                       \newcommand{\ibD}{{\boldsymbol{D}}}

\newcommand{\ibh}{{\boldsymbol{h}}}    \newcommand{\ibH}{{\boldsymbol{H}}}

\newcommand{\ibI}{{\boldsymbol{I}}}
\newcommand{\ibJ}{{\boldsymbol{J}}}
\newcommand{\ibla}{{\boldsymbol{\lambda}}}
\newcommand{\ibe}{{\bfm e}}
\newcommand{\ibm}{{\bfm m}}
\newcommand{\ibn}{{\bfm n}}             \newcommand{\ibN}{{\boldsymbol{N}}}

\newcommand{\ibp}{{\boldsymbol{p}}}     \newcommand{\ibP}{{\boldsymbol{P}}}
\newcommand{\ibr}{{\boldsymbol{r}}}
\newcommand{\ibs}{{\boldsymbol{s}}}
\newcommand{\ibt}{{\bfm t}}
\newcommand{\ibu}{{\boldsymbol{u}}}
\newcommand{\ibv}{{\boldsymbol{v}}}
\newcommand{\ibw}{{\boldsymbol{w}}}
\newcommand{\ibx}{{\boldsymbol{x}}}
\newcommand{\ibxizga}{\ibx_i, \ibz_{(i)}, \bga}
\newcommand{\ibxizgat}{\ibx_i, \ibz_{(i)}, \bga^{(t)}}

\newcommand{\ibX}{{\boldsymbol{X}}}
\newcommand{\iby}{{\boldsymbol{y}}}     \newcommand{\ibY}{{\boldsymbol{Y}}}
\newcommand{\ibz}{{\boldsymbol{z}}} \newcommand{\ibzt}{\ibz^{(t)}}
\newcommand{\ibzi}{\ibz_{(i)}}
\newcommand{\ibziT}{\ibz_{(i)}^{\T}}

\newcommand{\ibZ}{{\boldsymbol{Z}}}
\newcommand{\ibT}{{\boldsymbol{T}}}
\newcommand{\ibW}{{\boldsymbol{W}}}

% 6. hat,  tilde,  bar

% 6.1    hat
\newcommand{\hb}{\hat{b}}   \newcommand{\hbB}{\boldsymbol{\hat{\rm B}}}
\newcommand{\hm}{\hat{m}}
\newcommand{\hp}{\hat{p}}
\newcommand{\hq}{\hat{q}}
\newcommand{\hR}{\hat{R}}

\newcommand{\hbr}{\hat{\br}}

\newcommand{\hth}{\hat{\theta}}   \newcommand{\hTh}{\hat{\Theta}}
\newcommand{\hbth}{\boldsymbol{\hth}}
\newcommand{\hvth}{\hat{\vartheta}}
\newcommand{\hmu}{\hat{\mu}}    \newcommand{\hbmu}{\boldsymbol{\hmu}}
\newcommand{\hsi}{\hat{\sigma}}   \newcommand{\hSi}{\hat{\Sigma}}   \newcommand{\hbSi}{\boldsymbol{\hSi}}
\newcommand{\hal}{\hat{\alpha}}   \newcommand{\hbe}{\hat{\beta}}     \newcommand{\hbbe}{\bfm{\hat \beta}}
\newcommand{\hbal}{\boldsymbol{\hal}}
\newcommand{\hga}{\hat{\gamma}}  \newcommand{\hbga}{\boldsymbol{\hga}}
\newcommand{\hpsi}{\hat{\psi}}
\newcommand{\hxi}{\hat{\xi}}
\newcommand{\hpi}{\hat{\pi}} \newcommand{\hbpi}{\boldsymbol{\hpi}}
\newcommand{\hla}{\hat{\lambda}}  \newcommand{\hbla}{\bfm{{\hat \lambda}}}
\newcommand{\hde}{\hat{\delta}}
\newcommand{\hvp}{\hat{\varphi}}    \newcommand{\hbvp}{\boldsymbol{\hvp}}
\newcommand{\hrho}{\hat{\rho}}
\newcommand{\hbrho}{\boldsymbol{\hrho}}
\newcommand{\hbphi}{\bfm{\hat \phi}}
\newcommand{\hphi}{\hat \phi}

% 6.2    wide hat
\newcommand{\whVar}{\widehat{\mbox{Var}}}
\newcommand{\whse}{\widehat{\mbox{se}}}
\newcommand{\whSe}{\widehat{\mbox{Se}}}
\newcommand{\whPr}{\widehat{\Pr}}

\newcommand{\whbH}{\widehat{\bH}}
\newcommand{\whbS}{\widehat{\bS}}
\newcommand{\whbSi}{\widehat{\bSi}}

% 6.3    tilde

\newcommand{\tp}{\tilde{p}}

\newcommand{\tD}{\tilde{D}}
\newcommand{\tQ}{\tilde{Q}}
\newcommand{\tx}{\tilde{x}}
\newcommand{\ty}{\tilde{y}}

\newcommand{\tth}{\tilde{\theta}}     \newcommand{\tbth}{{\bfm \tth}}
\newcommand{\tpi}{\tilde{\pi}}
\newcommand{\tmu}{\tilde{\mu}}
\newcommand{\tbe}{\tilde{\beta}}
\newcommand{\tga}{\tilde{\gamma}}
\newcommand{\tsi}{\tilde{\sigma}}     \newcommand{\tSi}{\tilde{\Sigma}}
\newcommand{\tpsi}{\tilde{\psi}}
\newcommand{\txi}{\tilde{\xi}}
\newcommand{\tvp}{\tilde{\varphi}}

% 6.4     Bar
\newcommand{\Ba}{\bar{a}}       \newcommand{\Bat}[1]{\Ba_{#1}^{(t)}}
\newcommand{\Bb}{\bar{b}}       \newcommand{\Bbt}[1]{\Bb_{#1}^{(t)}}
\newcommand{\Bd}{\bar{d}}       \newcommand{\Bdt}[1]{\Bd_{#1}^{(t)}}
\newcommand{\Bs}{\bar{s}}       \newcommand{\Bst}[1]{\Bs_{#1}^{(t)}}
\newcommand{\Bt}{\bar{t}}
\newcommand{\Bu}{\bar{u}}
\newcommand{\Bv}{\bar{v}}
\newcommand{\Bw}{\bar{w}}
\newcommand{\Bx}{\bar{x}}
\newcommand{\By}{\bar{y}}       \newcommand{\BY}{\bar{Y}}
\newcommand{\Bz}{\bar{z}}

\newcommand{\Bxi}{\bar{\xi}}
\newcommand{\Bvth}{\bar{\vth}}

\newcommand{\BVar}{\overline{\mbox{Var}}}

% 7. Greek Letter

\newcommand{\al}{\alpha}            \newcommand{\alt}{\alpha^{(t)}}
                                    \newcommand{\altI}{\alpha^{(t+1)}}
\newcommand{\be}{\beta}
\newcommand{\ga}{\gamma}            \newcommand{\Ga}{\Gamma}
\newcommand{\gat}{\gamma^{(t)}}
\newcommand{\de}{\delta}            \newcommand{\De}{\Delta}
\newcommand{\la}{\lambda}           \newcommand{\lat}{\la^{(t)}}
                                    \newcommand{\latI}{\la^{(t+1)}}
\newcommand{\La}{\Lambda}
\newcommand{\tht}{\theta^{(t)}}
\newcommand{\thx}{\theta_x}         \newcommand{\Thx}{\Theta_x}
\newcommand{\thy}{\theta_y}
\newcommand{\piit}{\pi_i^{(t)}}

\newcommand{\ka}{\kappa}
\newcommand{\om}{\omega}            \newcommand{\Om}{\Omega}

\newcommand{\ve}{\varepsilon}
\newcommand{\vp}{\varphi}
\newcommand{\vr}{\varrho}
\newcommand{\vth}{\vartheta}

% 8. Bold Greek Letter

\newcommand{\bxi}{{\bfm \xi}}              \newcommand{\bet}{{\bfm \eta}}
\newcommand{\bphi}{{\boldsymbol{\phi}}}    \newcommand{\bphit}{\bphi^{(t)}}

\newcommand{\bmu}{{\boldsymbol{\mu}}}      \newcommand{\bnu}{{\boldsymbol{\nu}}}
\newcommand{\bla}{{\boldsymbol{\lambda}}}  \newcommand{\bLa}{{\bfm \Lambda}}
\newcommand{\bSi}{{\bfm \Sigma}}
\newcommand{\bom}{{\bfm \omega}}   \newcommand{\bOm}{{\bfm \Omega}}
\newcommand{\bde}{{\bfm \delta}}   \newcommand{\bDe}{{\bfm \Delta}}
\newcommand{\bth}{{\boldsymbol{\theta}}} \newcommand{\btht}{\bth^{(t)}}
\newcommand{\bTh}{{\boldsymbol{\Theta}}}
\newcommand{\bsi}{{\bfm \sigma}}
\newcommand{\bal}{{\boldsymbol{\alpha}}}
\newcommand{\bbe}{{\boldsymbol{\beta}}}
\newcommand{\brho}{{\boldsymbol{\rho}}}

\newcommand{\bpsi}{{\bfm \psi}}          \newcommand{\bPsi}{{\bfm \Psi}}
\newcommand{\bpi}{{\boldsymbol{\pi}}}    \newcommand{\bpit}{\bpi^{(t)}}
\newcommand{\bga}{{\boldsymbol{\gamma}}} \newcommand{\bgat}{\bga^{(t)}}
\newcommand{\bvp}{{\bfm \varphi}}        \newcommand{\bvpt}{\bvp^{(t)}}

% 9. distribution

\newcommand{\Bernoulli}{\mbox{Bernoulli}}
\newcommand{\Binomial}{\mbox{Binomial}}
\newcommand{\BBB}{\mbox{Bernoulli}_2^{\rm bk}}
\newcommand{\BBinomial}{\mbox{BBinomial}}
\newcommand{\Categorical}{\mbox{Categorical}}
\newcommand{\CP}{\mbox{CP}}
\newcommand{\D}{\mbox{D}}
\newcommand{\Degenerate}{\mbox{Degenerate}}
\newcommand{\DExponential}{\mbox{DExponential}}
\newcommand{\Dirichlet}{\mbox{\rm Dirichlet}}
\newcommand{\DMultinomial}{\mbox{DMultinomial}}
\newcommand{\Exponential}{\mbox{Exponential}}
\newcommand{\FDiscrete}{\mbox{FDiscrete}}
\newcommand{\Finite}{\mbox{Finite}}
\newcommand{\GD}{\mbox{GD}}
\newcommand{\GDirichlet}{\mbox{\rm GDirichlet}}
\newcommand{\GLiouville}{\mbox{GLiouville}}
\newcommand{\GPoisson}{\mbox{GPoisson}}
\newcommand{\GP}{\mbox{GP}}  \newcommand{\GPI}{\GP^{\mbox{\scriptsize{(I)}}}} \newcommand{\GPII}{\GP^{\mbox{\scriptsize{(II)}}}}
\newcommand{\Hgeometric}{\mbox{Hgeometric}}
\newcommand{\HPP}{\mbox{HPP}}
\newcommand{\IBeta}{\mbox{IBeta}}
\newcommand{\Ichi}{\mbox{I}\raisebox{0.5ex}{$\chi$}}
\newcommand{\IGamma}{\mbox{IGamma}}
\newcommand{\IGaussian}{\mbox{IGaussian}}
\newcommand{\IWishart}{\mbox{IWishart}}
\newcommand{\Laplace}{\mbox{Laplace}}
\newcommand{\Liouville}{\mbox{Liouville}}
\newcommand{\Logistic}{\mbox{Logistic}}
\newcommand{\Lognormal}{\mbox{Lognormal}}
\newcommand{\mBeta}{\mbox{Beta}}
\newcommand{\mGamma}{\mbox{Gamma}}
\newcommand{\mGaI}{\mbox{Ga}}

\newcommand{\Multinomial}{\mbox{Multinomial}}
\newcommand{\NBinomial}{\mbox{NBinomial}}
\newcommand{\ND}{\mbox{ND}}
\newcommand{\NDirichlet}{\mbox{NDirichlet}}
\newcommand{\NHPP}{\mbox{NHPP}}
\newcommand{\NIW}{\mbox{NIW}}
\newcommand{\Pareto}{\mbox{Pareto}}
\newcommand{\Poisson}{\mbox{Poisson}}
\newcommand{\TBeta}{\mbox{TBeta}}
\newcommand{\Wishart}{\mbox{Wishart}}
\newcommand{\ZIP}{\mbox{ZIP}}
\newcommand{\ZTP}{\mbox{ZTP}}
\newcommand{\ZAP}{\mbox{ZAP}}
\newcommand{\VZIP}{\mbox{VZIP}}
\newcommand{\ZDP}{\mbox{ZDP}}

%\newcommand{\f}[1]{f_{_{\tiny\rm #1}}}  \newcommand{\hf}[1]{\hat{f}_{#1}}

% 10. mbox -- statistics

\newcommand{\rhor}{\rho^{^{\rm real}}}
\newcommand{\rhof}{\rho^{^{\tiny \rm fit}}}

\newcommand{\AIC}{\mbox{AIC}}
\newcommand{\BIC}{\mbox{BIC}}
\newcommand{\Corr}{\mbox{Corr}} \newcommand{\CorrR}{\Corr^{\small \rm{R}}} \newcommand{\CorrF}{\Corr^{\small \rm{fit}}}
\newcommand{\Cov}{\mbox{Cov}}
\newcommand{\sol}{\mbox{sol}}
\newcommand{\se}{\mbox{se}}

\newcommand{\SE}{\mbox{SE}}
\newcommand{\sgn}{{\rm sgn}}
\newcommand{\tr}{\mbox{$\,$tr$\,$}}
\newcommand{\rank}{\mbox{rank}\,}
\newcommand{\Var}{\mbox{Var}}
\newcommand{\Skew}{\mbox{Skew}}
\newcommand{\MSE}{\mbox{MSE}}
\newcommand{\CV}{\mbox{CV}}
\newcommand{\median}{\mbox{median}}

\newcommand{\logit}{\mbox{logit}}
\newcommand{\diag}{\mbox{diag}}
\newcommand{\data}{\mbox{data}}
\newcommand{\KL}{\mbox{KL}}
\newcommand{\IG}{\mbox{IG}}
\newcommand{\ISE}{\mbox{ISE}}

\newcommand{\qand}{\quad \mbox{and} \quad}
\newcommand{\qqand}{\quad \qand \quad}
\newcommand{\qas}{\quad \mbox{as} \quad}
\newcommand{\qag}{\quad \mbox{against} \quad}
\newcommand{\qfor}{\quad \mbox{for} \quad}
\newcommand{\qor}{\quad \mbox{or} \quad}
\newcommand{\qve}{\quad \mbox{versus} \quad}
\newcommand{\qwh}{\quad \mbox{where} \quad}
\newcommand{\qwi}{\quad \mbox{with} \quad}
\newcommand{\col}{\mbox{: }}
\newcommand{\RE}{\mbox{RE}}
\newcommand{\yes}{\mbox{yes}}
\newcommand{\No}{\mbox{no}}
\newcommand{\DPP}{\mbox{DPP}}

\newcommand{\rd}{\,\mbox{d}}
\newcommand{\w}{\mbox{w}}
\renewcommand{\ge}{\geqslant}
\renewcommand{\le}{\leqslant}
\newcommand{\Zp}{Z\hspace{0.02cm}'}

% 11. Simplification

\newcommand{\I}{{\rm I}}
\newcommand{\II}{I\hspace*{-0.4mm}I}
\newcommand{\III}{I\!I\!I}
\newcommand{\IR}{{I\!\! R}}
\newcommand{\IV}{I$\!$V}
\newcommand{\et}{{\it et al}.}
\newcommand{\Et}{{\it et al}.\,}

\newcommand{\yikH}{y_{ik}^{\rm H}}
\newcommand{\ybH}{\bar{y}^{\rm H}}

% 12. Sign

\newcommand{\sd}{\stackrel{{\rm d}}{=}}
\newcommand{\sdt}{$ {\small $\sd$} $}
\newcommand{\heq}{\;\hat{=}\;}
\newcommand{\teq}{\triangleq}
\newcommand{\iid}{\stackrel{{\rm iid}}{\sim}}
\newcommand{\tiid}{i.i.d.$\hspace*{0.08cm}$}
\newcommand{\ind}{\stackrel{{\rm ind}}{\sim}}
\newcommand{\dsim}{\stackrel{.}{\sim}}
\newcommand{\dis}{\displaystyle}
\newcommand{\tex}{\textstyle}
\newcommand{\cf}{cf.$\hspace*{0.1cm}$}
\newcommand{\trv}{r.v.$\hspace*{0.08cm}$}

\newcommand{\T}{\!\top\!}
\newcommand{\na}{\nabla}
\newcommand{\noi}{\noindent}
\newcommand{\ra}{\rightarrow}
\newcommand{\pr}{\propto}
\newcommand{\eq}{\equiv}
\newcommand{\pa}{\partial}
\newcommand{\ol}{\overline}
\newcommand{\non}{\nonumber}
\newcommand{\ap}{\approx}
\newcommand{\Bot}{\;\bot\;}
\newcommand{\inde}{{\Bot\!\!\!\!\!\!\!\Bot}}
\newcommand{\btu}{\bigtriangleup}

% 13. vskip

\newcommand{\vs}{\vspace*{-0.25cm}}
\newcommand{\vkl}{\vskip 0.10in}
\newcommand{\vkL}{\vskip 0.15in}
\newcommand{\vkU}{\vskip 0.30in}

% 14. environment

\newcommand{\namelistlabel}[1]{\mbox{#1}\hfil}
\newenvironment{namelist}[1]{%
\begin{list}{}
       {\let \makelabel \namelistlabel
        \settowidth{\labelwidth}{#1}
        \setlength{\leftmargin}{1.1\labelwidth}   }
        }{%
\end{list} }

% 15. other
\def\bds{\begin{description} \itemsep=-\parsep \itemindent=-0.7 cm}
\def\eds{\end{description}}
\def\i{\item}

% Liu Yin
%\newcommand{\hphi}{\hat{\phi}}
\newcommand{\tphi}{\tilde{\phi}}
\newcommand{\tla}{\tilde{\lambda}}
\newcommand{\MP}{\mbox{MP}}
\newcommand{\twocc}[2]{\bigg\{\!\!\begin{array}{c} #1 \\ #2  \end{array} \!\!\bigg\}}
\newcommand{\bibx}{{\boldsymbol{\bar{x}}}}
\newcommand{\tbmu}{{\boldsymbol{\tilde{\mu}}}}
\newcommand{\gD}{\mbox{\rm gD}}
\newcommand{\HDirichlet}{\mbox{\rm HDirichlet}}
\newcommand{\hbp}{\bfm {\hat p}}

% Lixunjian

\newcommand{\tal}{\tilde{\al}}
\newcommand{\Bpi}{\bar{\pi}}
\newcommand{\Bal}{\bar{\al}}
\newcommand{\Bbe}{\bar{\be}}
\newcommand{\p}[1]{p_{\rm #1}}
\newcommand{\s}[1]{s_{\rm #1}}
\newcommand{\GA}{\mbox{GA}}
\newcommand{\NR}{\mbox{NR}}
\newcommand{\zed}{\mbox{zeta}}

\newcommand{\Casesthree}[6]
{\left\{\!\!\begin{array}{ll} #1 & #2  \\ [2mm]
                              #3 & #4  \\ [2mm]
                              #5 & #6
\end{array} \right. }

\newcommand{\mif}{\mbox{if } \ }

\newcommand{\tha}{\theta}           \newcommand{\Tha}{\Theta}
\newcommand{\ths}{\theta^*}
\newcommand{\thtI}{\theta^{(t+1)}}

% Zhao Yuanfan

\newcommand{\email}[1]{\href{mailto:#1}{\nolinkurl{#1}}}

% Paper 1: GeGa

\newcommand{\mpr}{\mbox{Pr}}  % = \Pr
\newcommand{\e}{{\rm e}}
\newcommand{\Kur}{\mbox{Kurt}}
\newcommand{\GeGa}{\mbox{Ge-Ga}(\al,\mu,\bla)}
\newcommand{\IGaGa}{\mbox{IGa-Ga}(\al,\mu, \la)}
\newcommand{\IGauGa}{\mbox{IGau-Ga}(\al, \mu, \la)}
\newcommand{\RIGauGa}{\mbox{RIGau-Ga}(\al, \mu, \la)}
\newcommand{\PaGa}{\mbox{Pa-Ga}(\al, \mu, \la)}
\newcommand{\LN}{\mbox{LN}}
\newcommand{\bthtt}{\bth^{(t+1)}}   \newcommand{\bthtI}{\bth^{(t+1)}}
\newcommand{\bbet}{\bbe^{(t)}}
\newcommand{\bbett}{\bbe^{(t+1)}}   \newcommand{\bbetI}{\bbe^{(t+1)}}
\newcommand{\mut}{\mu^{(t)}}
\newcommand{\mutt}{\mu^{(t+1)}}     \newcommand{\mutI}{\mu^{(t+1)}}
\newcommand{\GIG}{\mbox{GIG}}
\newcommand{\GIGau}{\mbox{GIGau}}
\newcommand{\IGa}{\mbox{IGa}}
\newcommand{\RIG}{\mbox{RIG}}
\newcommand{\RIGau}{\mbox{RIGau}}
\newcommand{\Weibull}{\mbox{Weibull}}
\newcommand{\Se}{\mbox{Se}}
\newcommand{\constant}{\mbox{constant}}

% Paper 2: MVGA
\newcommand{\GeGad}{\mbox{Ge-Ga}_d(\al,\bmu,\bla)}
\newcommand{\IGaGad}{\mbox{IGa-Ga}_d(\al,\bmu, \la)}
\newcommand{\LSGIGauGa}{\mbox{LSGIGau-Ga}_d(\al, \bmu, \la)}
\newcommand{\LNGa}{\mbox{LN-Ga}_d(\al,\bmu,\la)}
\newcommand{\tsf}[1]{\textsf{#1}}

% Paper 3: NP-Ga
\newcommand{\NPGa}{\mbox{NP-Ga}_d(\al, \bmu, \ibp)}
\newcommand{\ibg}{\boldsymbol{g}}   \newcommand{\ibgt}[1]{\ibg_{#1}^{(t)}}
\newcommand{\ptI}[1]{p^{(t+1)}_{#1}}
\newcommand{\bvth}{\boldsymbol{\vth}}
\newcommand{\mcL}{\mathcal{L}}

% Paper 4: NPMN

\newcommand{\GeN}{\mbox{Ge-N}_d(\bmu, \bSi, \bla)}
\newcommand{\Kurt}{\mbox{Kurt}}
\newcommand{\At}[1]{\bA_{#1}^{(t)}}
\newcommand{\bmut}{\bmu^{(t)}}
\newcommand{\bmutI}{\bmu^{(t+1)}}
\newcommand{\bSit}{\bSi^{(t)}}  \newcommand{\bSitI}{\bSi^{(t+1)}}
\newcommand{\bBtI}{\bB^{(t+1)}}
\newcommand{\bgatI}{\bga^{(t+1)}}
\newcommand{\btau}{\boldsymbol{\tau}}

\newcommand{\GaN}{\mbox{Ga-N}_d(\bmu, \bSi, \la)}
\newcommand{\IGaN}{\mbox{IGa-N}_d(\bmu, \bSi, \la)}
\newcommand{\IGauN}{\mbox{IGau-N}_d(\bmu, \bSi, \la)}
\newcommand{\RIGauN}{\mbox{RIGau-N}_d(\bmu, \bSi, \la)}

\newcommand{\NPMN}{\mbox{NPM-N}_d(\bmu, \bSi, \ibp)}
\newcommand{\bvptI}{\bvp^{(t+1)}}
\newcommand{\Loglikelihood}{\textsf{Log-likelihood}}
\newcommand{\Rank}{\textsf{Rank}}
\newcommand{\MAE}{\textsf{MAE}}

% Paper 5: BSB

\newcommand{\Bernstein}{\mbox{Bernstein}}
\newcommand{\BetaBino}{\mbox{Beta-Bino}}
\newcommand{\BernBino}{\mbox{Bern-Bino}}
\newcommand{\blat}{\bla^{(t)}}
\newcommand{\blatI}{\bla^{(t+1)}}
\newcommand{\ibV}{\boldsymbol{V}}
\newcommand{\ibbt}[1]{\ibb^{(t)}_{#1}}
\newcommand{\bBt}{\bB^{(t)}}
\newcommand{\hblas}{\hbla^*}
\newcommand{\cL}{{\cal L}}
\newcommand{\tbla}{\boldsymbol{\tla}}
\newcommand{\Bla}{\bar{\la}}
\newcommand{\tod}{\overset{\text{D}}{\rightarrow}}

\voffset -0.5truecm \hoffset -1.5truecm

%!TEX root = BSB.tex

\begin{flushleft}
{\Large \bf A new Bernstein--binomial model for fitting finite discrete data with over--dispersion: Likelihood-based and Bayesian approaches}
\end{flushleft}

\smallskip
\begin{flushleft}
{\bf Yuan-Fan ZHAO$^{{\rm a}, \dag, *}$},   \ \
{\bf Yikai GUO$^{{\rm a}, \dag}$},   \ \
{\bf Xun-Jian LI$^{\rm b}$},   \ \  
{\bf Man-Lai TANG$^{\rm c}$}   \ \  and \ \
{\bf Guo-Liang TIAN$^{{\rm a}, *}$}  \ \

\vkL
{\small \it $^{\rm a}$Department of Statistics and Data Science,
                      Southern University of Science and Technology, \\ Shenzhen 518055, Guangdong Province, P. R. China} \\
\vkL
{\small \it $^{\rm b}$Department of Biostatistics, University of California, Los Angeles, Los Angeles, CA 90095, USA}

\vkL
{\small \it $^{\rm c}$Centre of Data Innovation Research, Department of Physics, Astronomy and Mathematics, School of Physics, Engineering and Computer Science, University of Hertfordshire, College Lane, Hatfield, UK}

\vkL
{\small  $^{\dag}$Equally contributed} \\
{\small  $^*$Co-corresponding authors' Emails: \tsf{\email{12431116@mail.sustech.edu.cn} (YF ZHAO)} and \tsf{\email{tiangl@sustech.edu.cn} (GL TIAN) }} \\
{\small \textsf{4 OCT 2026 ZYF; 4 OCT 2026 Gary}}
\end{flushleft}

\baselineskip 0.25in \vskip 0.05in \noi {\bf Abstract:}  Finite discrete count data are ubiquitous in fields such as biomedicine and social sciences and these data frequently exhibit over--dispersion, rendering the binomial distribution inadequate for modeling. Although the beta--binomial distribution mitigates this by assigning a subjective beta prior to the success probability, its reliance on a pre-specified parametric shape may result in biased representations of prior information, failing to capture the complex and heterogeneous characteristics of real--world data. To address these limitations, this paper proposes a novel Bernstein--binomial model. By leveraging the uniform approximation properties of Bernstein distribution, we construct a highly flexible and general prior framework for the success probability in binomial distribition, which dynamically adapts to complex data structures, avoiding the subjectivity and restrictions associated with conventional unimodal priors. A comprehensive theoretical framework for both frequentist and Bayesian inferences is established. Furthermore, the model is extended to a regression setting to account for covariate effects. Extensive simulation studies and real--world data applications confirm that the proposed Bernstein--binomial model significantly outperforms existing models. It offers superior fitting accuracy, robustness, and out-of-sample predictive performance when modeling over--dispersed counts with complex latent prior structures.

\vkL \noi {\bf Keywords:}   Bayesian inference, Bernstein polynomial, Binomial distribution, MM algorithm, Regression model.

\baselineskip 0.30in
\setcounter{equation}{0} 

%!TEX root = BSB.tex

\section{$\!\!\!\!\!\!\!$. Introduction}    % Section 1

\subsection{Literature review}      % Section 1.1

Data defined on the finite discrete set $\bbN_{0:m} \teq \{0, 1, \dots, m\}$ widely exists in empirical research (\cite{MN2019}), and its essence is the counting observation constrained by the upper bound $m$. This kind of data is universal in biomedical fields (number of gene mutation sites detected), industrial engineering (number of product defects under fixed sampling) and social sciences (frequency statistics of discrete scoring options).

A discrete {\it random variable} (r.v.) $X$ is said to follow a binomial distribution, denoted by $X \sim \Binomial(m, p)$, if its \textit{probability mass function} (pmf) is
$$
   \Binomial(x|m,p) = \binom{m}{x} p^x (1-p)^{m-x}, \quad x \in \bbN_{0:m},
$$
where mean $E(X)=mp$ and variance $\Var(X)=mp(1-p)=E(X)[1-E(X)/m]$, which is a quadratic function of $E(X)$; i.e., $\Var(X)$ is completely determined by $E(X)$. However, the actual finite discrete data often violate their ideal assumptions: First, the phenomenon of over--dispersion (the observation variance is significantly greater than $mp(1-p)$) is widespread in the actual data (\cite{Williams1982} and \cite{Moore1987}). Secondly, the homogeneous $p$ hypothesis is difficult to adapt to heterogeneous groups (\cite{Crowder1978} and \cite{HK1979}).

In order to address the issue that binomial distribution cannot fit finite discrete data with over--dispersion, scholars propose a  \textit{beta--binomial} (Beta-Bino) distribution from the Bayesian viewpoint (\cite{Kleinman1973}), where the success probability $p$ is treated as a \trv (\cite{Skellam1948}) and a $\mBeta(\al, \be)$ distribution is assigned as a prior of $p$, resulting in the pmf of $Y \sim \BetaBino(m, \al, \be)$ as
\begin{eqnarray} \label{zyffour1.add1}
   \BetaBino(y|m, \al, \be) = \binom{m}{y} \frac{B(\al+y, \be + m-y)} {B(\al,\be)}, \quad y \in \bbN_{0:m},
\end{eqnarray}
where $B(a,b)$ represents the beta function, the mean $E(Y)=m\al/(\al + \be) \teq m\mu$ and the variance
$$
   \Var(Y) = m\mu(1-\mu) \left(1 + \frac{m-1}{\al + \be + 1} \right)
$$
explicitly contains the over--dispersion correction term.

Main reason for selecting beta distribution as the prior distribution of $p$ is that beta and binomial distributions are conjugate, leading that beta--binomial pmf has explicit expression and corresponding statistical calculations are easy to conduct. In fact, the prior of $p$ can select any continuous distribution defined on the interval $[0,1]$ (\cite{DY1979}), such as logit-normal distribution, Kumaraswamy distribution, and a set of distributions proposed by \textcite{LT2025}. The arbitrary choice of prior distribution makes the model fitting too subjective, and it is impossible to judge the true prior of the data in the actual model fitting. Therefore, in this paper, based on Weierstrass approximation theorem, we propose to use Bernstein distribution as a general prior distribution of $p$ in binomial distribution, resulting in a new Bernstein–-binomial model which can be employed to fit finite discrete data with over--dispersion.

The rest of the paper is organized as follows. In Section 1.2, we will introduce Weierstrass approximation theorem and Bernstein distribution. In Section 2, we propose Bernstein--binomial distribution for the first time and use \textit{minorization--maximization} (MM, \cite{HL2004}) algorithm, \textit{data augmentation} (DA, \cite{TW1987}) algorithm and other methods to conduct statistical inferences. The regression model of Bernstein--binomial distribution is investigated in Section 2.5. Bayesian inferences of our proposed distribution is provided in Section 3. Some numerical experiments are presented in Section 4, which can prove the reliability of our model and the accuracy of corresponding algorithms. In Section 5, we will apply this new model to real data analysis and compare it with some traditional models. The conclusion and discussion of the results are stated in Section 6. Finally, the technical details and derivation is put in the appendix.

\subsection{Weierstrass approximation theorem and Bernstein \\ distribution}     % Section 1.2

Given a non-negative integer $K$, Bernstein polynomials of order $K$ (\cite{Petrone1999}, \cite{Ghosal2001} and \cite{PW2002}) is a set of $K+1$ polynomials $\{B_{k,K}(\tau)\col k \in \bbN_{0:K}\}$, where the individual polynomial
\begin{eqnarray} \label{zyffour1.1}
   B_{k,K}(\tau) \teq  \binom{K}{k} \tau^k (1-\tau)^{K-k}
   = \frac{1}{K+1} \cdot \mBeta(\tau| k+1, K-k+1), \quad \tau \in [0,1],
\end{eqnarray}
is sometimes called as the $k$-th Bernstein basis function of order/degree $K$, here $\mBeta(\cdot|a,b)$ denotes the \textit{probability density function} (pdf) of the $\mBeta(a, b)$ distribution. %Key Properties of Bernstein polynomials include

Bernstein polynomials are particularly useful in approximation theory.
Given a continuous function $f(\tau)$ on the interval $[0,1]$, the $K$-order \textit{Bernstein polynomial approximation} (BPA) is defined as
\begin{eqnarray}  \label{zyffour1.2}
   B_K(\tau|f) &\teq & \sum_{k=0}^K f(k/K) \cdot B_{k,K}(\tau)
   \sr{(\ref{zyffour1.1})}  \sum_{k=0}^K  \frac{ f(k/K)}{K+1} \cdot \mBeta(\tau|k+1, K-k+1) \non \\ [2mm]
   &=& \sum_{k=0}^K w_k \cdot \mBeta(\tau|k+1, K-k+1),  \quad \tau \in [0,1],
\end{eqnarray}
where $w_k \teq f(k/K)/(K+1)$. The formula (\ref{zyffour1.2}) indicates that the $K$-order BPA can be expressed as a linear combination of $K+1$ beta densities.

\begin{theorem} \label{thm1} {\rm  \textsf{(Weierstrass approximation theorem)}.  For any continuous function $f(\tau)$ defined on $[0,1]$, the $K$-order BPA $B_K(\tau |f)$ uniformly converges to $f(\tau)$; that is,
$$
   \lim_{K\to \infty} \sup_{\tau\in[0,1]} |B_K(\tau|f) - f(\tau) | = 0. \eqno{\|}
$$
}\end{theorem}

\subsubsection{Bernstein distribution: A generalization of beta distribution}     % Section 1.2.1

Consider a continuous \trv $X^* \in [0,1]$, if its pdf $f^*(\cdot)$ is a continuous function defined on $[0,1]$, then Theorem \ref{thm1} indicates that $f^*(\cdot)$ can be uniformly approximated by
\begin{eqnarray}  \label{zyffour1.3}
   \Bernstein(\tau|K, \bla) \teq \sum_{k=0}^K \la_k \cdot \mBeta(\tau|k+1,K-k+1), \quad \tau \in [0, 1].
\end{eqnarray}
The formula (\ref{zyffour1.3}) defines the pdf of a so-called $\Bernstein(K, \bla)$ distribution, where $K$ is a known non-negative integer (or a predetermined constant), and  $\bla = (\la_0,\la_1,\ldots,\la_K)^{\T}$ is an unknown parameter vector of weights satisfying $\bla \in \bbT_{K+1}$, where
\begin{eqnarray}   \label{zyffour1.4}
   \bbT_n \teq \left\{\ibx=(x_1, \dots, x_n)^{\T}\col 0 \le x_i \le 1, \; i \in \bbN_{1:n}, \; \ibx^{\T}\1_n = 1 \right\}
\end{eqnarray}
is the $n$-dimensional {\it closed} hyperplane, and $\1_n$ is the $n$-vector with each element being 1. For convenience, we define the $n$-dimensional {\it open} hyperplane as
\begin{eqnarray}   \label{zyffour1.5}
   \bbT_n^{\rm open} \teq \left\{\ibx=(x_1, \dots, x_n)^{\T}\col 0 < x_i < 1, \; i \in \bbN_{1:n}, \; \ibx^{\T}\1_n = 1 \right\}.
\end{eqnarray}

We illustrate that the parameter vector $\bla$ in the Bernstein distribution is identifiable as follows: Because the Bernstein basis functions $\{B_{k,K}(\tau) \col k \in \bbN_{0:K}\}$ are linearly independent on the interval $[0,1]$ and forms the basis of a $(K+1)$-dimensional polynomial space. That is, any polynomial with a degree not exceeding $K$ can be uniquely represented as a linear combination of these basis functions. Therefore, the beta distributions  $\{\mBeta(\tau|k+1,K-k+1) \ | \ k \in \bbN_{0:K}\}$ are also linearly independent, ensuring the uniqueness and parameter identifiability of the Bernstein distribution as a finite discrete mixture of beta distributions.

Figure 1 shows the approximate effect of Bernstein pdf on some common continuous distributions defined on the interval $[0,1]$. We found that with the increase of $K$, the Bernstein pdf gradually approaches the target pdf, and finally achieves a completely approximate effect (\cite{BCC2002}, \cite{Kakizawa2004} and \cite{TS2014}).

Based on the above theorems and conclusions, we can take Bernstein distribution as the general prior distribution of the success probability $p$ in the binomial distribution, which can not only expand the constraint prior of the original binomial distribution, adapt to more complex and changeable real data, but also avoid the subjectivity and randomness of the selection of prior distribution, there is no need to preset the specific form.

\begin{figure}[H]
    \subfigure[$\;$ $\mbox{Logit-N}(-1,1)$]
    {
        \centering
        \includegraphics[width=0.5\linewidth]{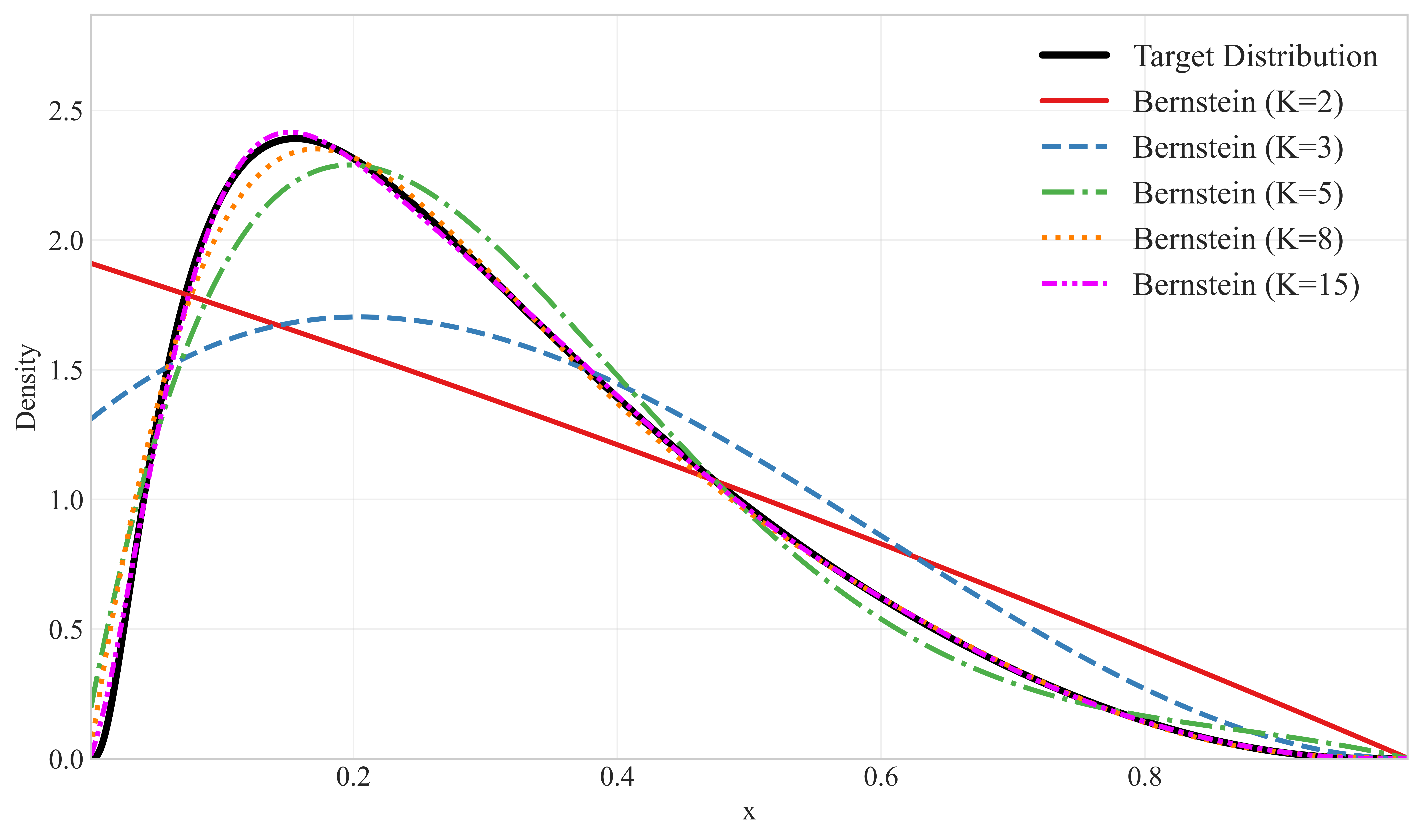}
    }
    \subfigure[$\;$ $\mBeta(2,5)$]
    {
        \centering
        \includegraphics[width=0.5\linewidth]{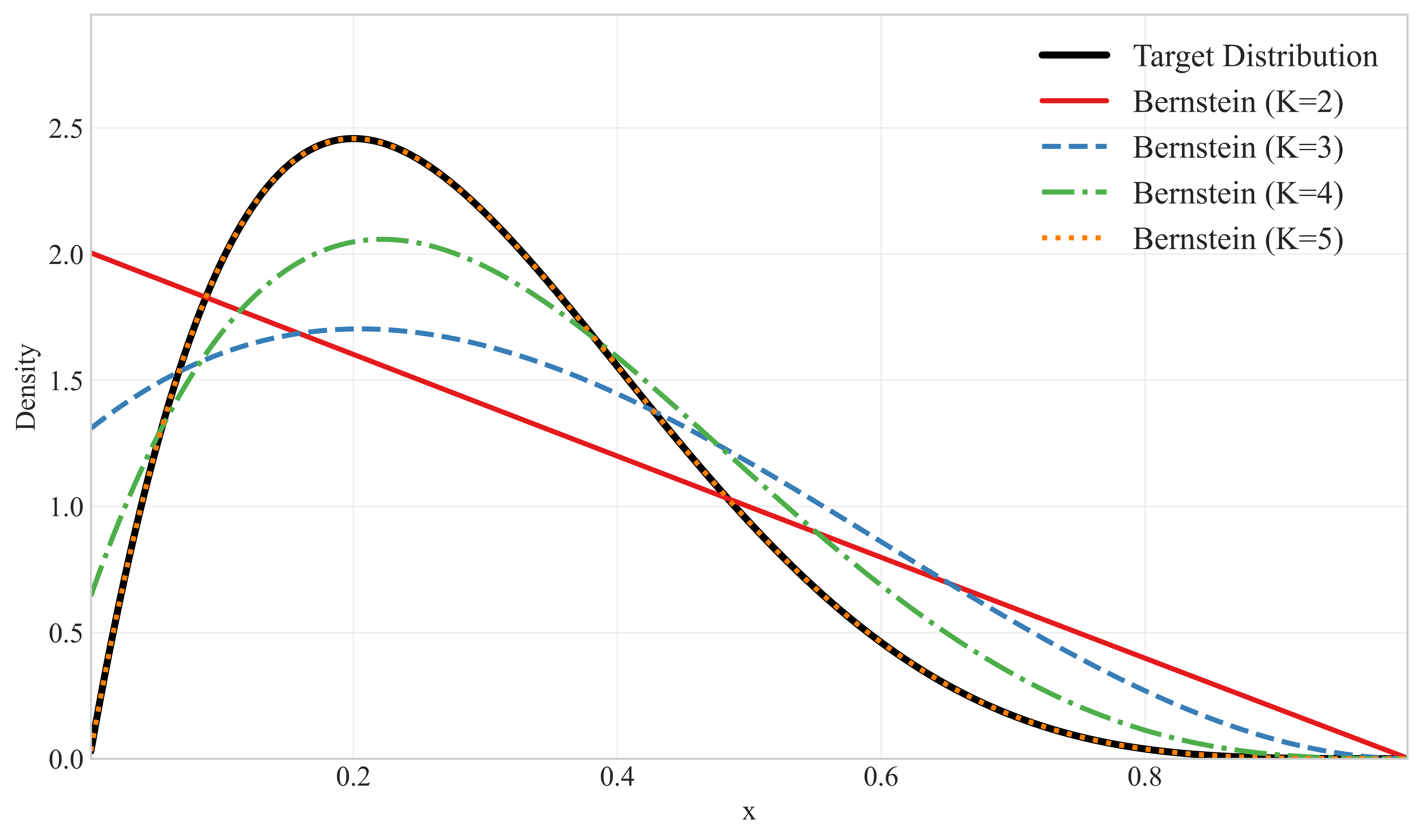}
    }
    \subfigure[$\;$ $\mbox{Kumaraswamy}(2,3)$]
    {
        \centering
        \includegraphics[width=0.5\linewidth]{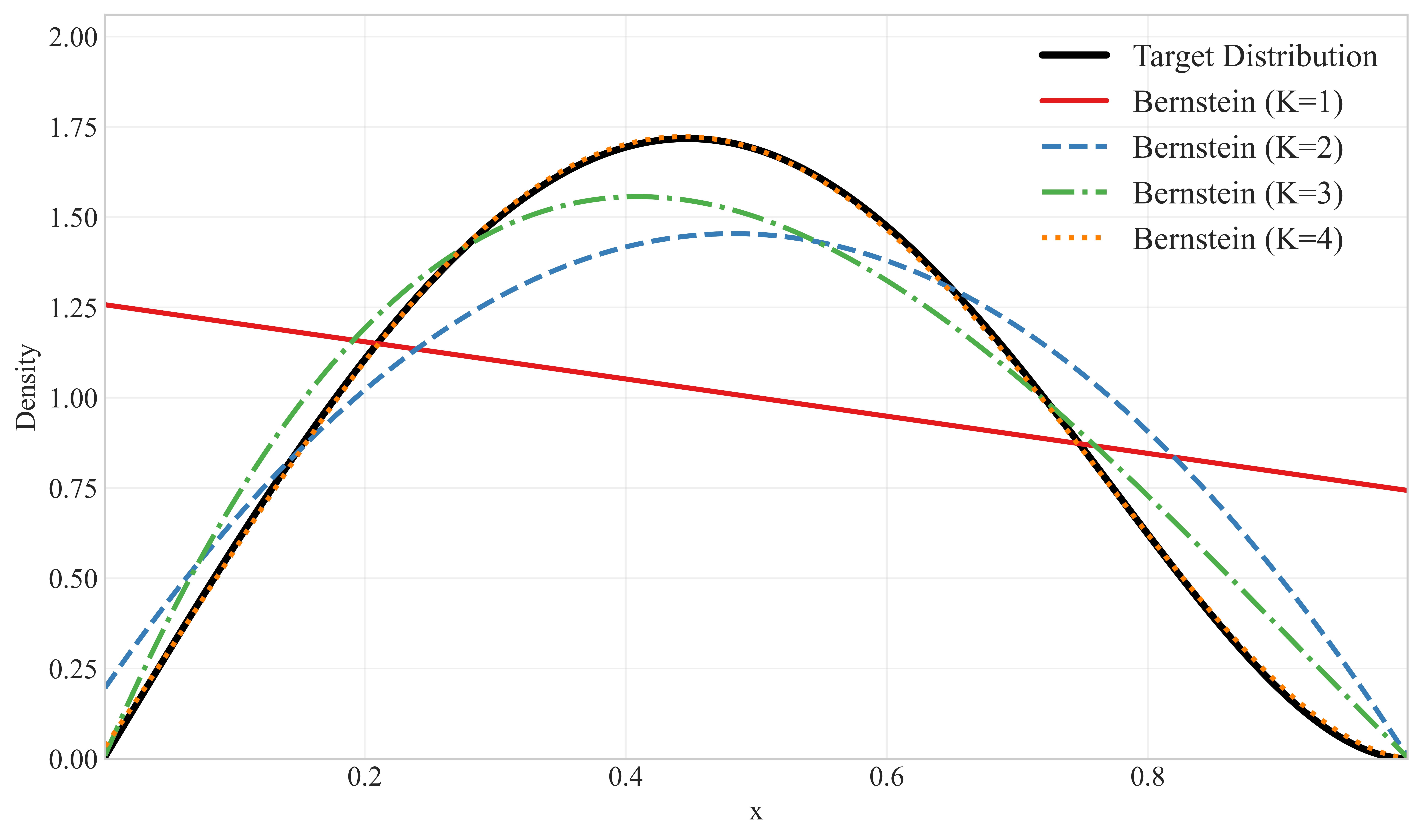}
    }
    \subfigure[$\;$ $\mbox{Triangular}(0.7)$]
    {
        \centering
        \includegraphics[width=0.5\linewidth]{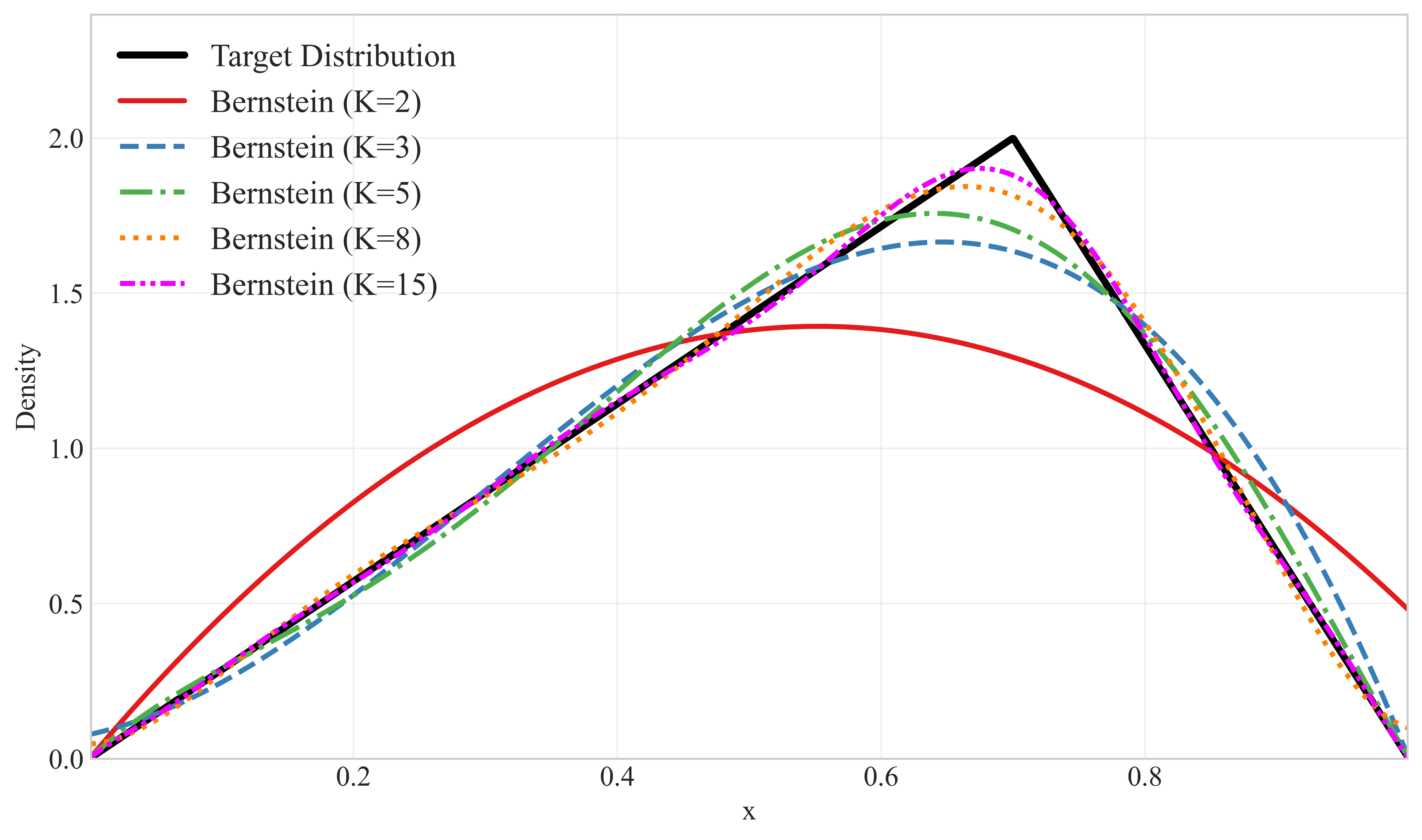}
    }
    \caption{$\;$ As $K$ increases, Bernstein densities $\Bernstein(\tau|K, \bla)$ approximate various target densities defined on $[0,1]$: (a) logit-normal distribution Logit-N(-1, 1); (b) Beta distribution Beta(2, 5); (c) Kumaraswamy distribution Kumaraswamy(2,3); (d) Triangular distribution Triangular(0.7). }
\end{figure} 
%!TEX root = BSB.tex

\section{$\!\!\!\!\!\!\!$. Bernstein--binomial distribution}
% Section 2

\subsection{Definition and basic properties}   % Section 2.1

Let $\tau$ be a continuous \trv taking values on the unit interval $[0,1]$. If a discrete \trv $X$ has the following \textit{mixture representation} (MR):
\begin{eqnarray}   \label{eqn2.1}
   \tau \sim \Bernstein(K, \bla) \qand X|\tau \sim \Binomial(m,\tau),
\end{eqnarray}
then $X$ is said to follow the \textit{Bernstein--binomial} (Bern-Bino) distribution, denoted by $X \sim \BernBino(m, K, \bla)$, where $\{m, K\}$ are two known constants and $\bla = (\la_0, \la_1, \dots, \la_K)^{\T} \in \bbT_{K+1}$ is an unknown weights vector. Its pmf is derived as
\begin{eqnarray}  \label{eqn2.2}
   \BernBino(x|m, K, \bla) = \binom{m}{x} \sum_{k=0}^K  a_k(x) \la_k = \binom{m}{x} [\iba(x)]^{\T} \bla, \quad x \in \bbN_{0:m}, \quad
\end{eqnarray}
where $\iba(x) =(a_0(x), a_1(x), \dots, a_K(x))^{\T}$ and
\begin{eqnarray*}
   a_k(x) \teq \frac{B(k+1+x, K-k+1+m-x)}{B(k+1,K-k+1)}, \quad k \in \bbN_{0:K}.
\end{eqnarray*}

From the MR (\ref{eqn2.1}), on the one hand, we can obtain
\begin{eqnarray*}
   E(X) &=& \frac{m}{K+2} \sum_{k=0}^K (k+1)\la_k \qand   \\ [2mm]
   \Var(X) &=& m \left\{ \frac{\sum_{k=0}^K (k+1)\la_k}{K+2} - \left[ \frac{\sum_{k=0}^K (k+1)\la_k}{K+2} \right]^2 \right\}  \\ [2mm]
   & & +\; \frac{m(m-1)}{(K+2)^2(K+3)} \sum_{k=0}^K (k+1)(K-k+1) \la_k.
\end{eqnarray*}
On the other hand, we have the \textit{stochastic representation} (SR) of $\tau \sim \Bernstein(K, \bla)$ as
\begin{eqnarray} \label{zyffour2.3}
   \tau \sd Z_0 Y_0 + Z_1 Y_1 + \cdots + Z_K Y_K = \ibZ^{\T} \ibY,
\end{eqnarray}
where $\ibZ = (Z_0,Z_1,\ldots,Z_K)^{\T} \sim \Multinomial(1,\bla)$, $\{Y_k\}_{k=0}^K \ind \mBeta(k+1,K-k+1)$, and $\ibZ \inde \ibY = (Y_0, Y_1,\ldots, Y_K)^{\T}$. The SR (\ref{zyffour2.3}) can be employed to generate random samples from the $\Bernstein(K, \bla)$ distribution and to make Bayesian inferences in the DA algorithm (see \S 3).

\begin{remark} \label{rem1} {\rm (\textsf{Alternative expression of pmf for the Bern-Bino distribution}). Based on (\ref{zyffour1.add1}), we can rewrite (\ref{eqn2.2}) as a mixture distribution of $(K+1)$ beta--binomial distributions:
\begin{eqnarray}  \label{eqn2.TGL4}
   \BernBino(x|m, K, \bla) \sr{(\ref{eqn2.2})} \sum_{k=0}^K  \la_k \cdot \BetaBino(x|m, k+1, K-k+1), \quad x \in \bbN_{0:m}.
\end{eqnarray}
Hence, similar to (\ref{zyffour2.3}), we can obtain the SR of $X \sim \BernBino(m, K, \bla)$ as
\begin{eqnarray} \label{zyffour2.TGL5}
   X \sd Z_0 V_0 + Z_1 V_1 + \cdots + Z_K V_K = \ibZ^{\T} \ibV,
\end{eqnarray}
where $\ibZ = (Z_0,Z_1,\ldots,Z_K)^{\T} \sim \Multinomial(1,\bla)$, $\{V_k\}_{k=0}^K \ind \BetaBino(m, k+1, K -k+1)$, and $\ibZ \inde \ibV = (V_0, V_1,\ldots, V_K)^{\T}$. The SR (\ref{zyffour2.TGL5}) can be used to alternatively generate random samples from the $\BernBino(m, K, \bla)$ distribution parallel to the MR (\ref{eqn2.1}). \hfill $\|$
}\end{remark}

\subsection{MLEs of parameters via an MM algorithm}  % Section 2.2

Let $\{X_i\}_{i=1}^n \iid \BernBino(m, K, \bla)$ and $\Y{obs} = \{x_i\}_{i=1}^n$ denote the observations, then the log-likelihood function of $\bla$ is
\begin{eqnarray} \label{zyffour2.4}
   \ell(\bla|\Y{obs}) \sr{(\ref{eqn2.2})} \sum_{i=1}^n \log\binom{m}{x_i} + \sum_{i=1}^n \log \left( \sum_{k=0}^K a_{ik} \la_k \right) = \constant + \sum_{i=1}^n \log (\iba_i^{\T}\bla),
\end{eqnarray}
where $\iba_i =(a_{i0}, a_{i1}, \dots, a_{iK})^{\T}$,
\begin{eqnarray}   \label{zyffour2.5}
   a_{ik} = \frac{B(k+1+x_i,K-k+1+m-x_i)}{B(k+1,K-k+1)}, \quad i \in \bbN_{1:n}, \; k\in \bbN_{0:K},
\end{eqnarray}
are positive constants free from $\bla \in \bbT_{K+1}$. The aim of this subsection is to calculate the MLEs of $\bla$ through an MM algorithm.

To this end, we first note the discrete version of Jensen's inequality
$$
    \vp\left(\sum_{k=0}^K b_k z_k \right) \ge \sum_{k=0}^K b_k \vp(z_k),
$$
for any concave function $\vp(\cdot)$, where $\ibb=(b_0, b_1, \dots, b_K)^{\T} \in \bbT_{K+1}$. For a fixed $i$, by setting $\vp(\cdot) = \log(\cdot)$, we then have the following inequality
\begin{eqnarray} \label{zyffour2.6}
   \log (\iba_i^{\T}\bla) \ge \constant + \sum_{k=0}^K \bt{ik} \log(\la_k),
\end{eqnarray}
where the weights
\begin{eqnarray} \label{zyffour2.7}
   \bt{ik} \teq \frac{a_{ik}\lat_k}{\sum_{l=0}^K a_{il}\lat_l} = \frac{a_{ik}\lat_k}{\iba_i^{\T}\blat}, \quad k\in \bbN_{0:K},
\end{eqnarray}
and $\blat =(\lat_0, \lat_1, \dots, \lat_K)^{\T}$ represents the $t$-th approximation of the MLE $\hbla$. By combining (\ref{zyffour2.4}) with (\ref{zyffour2.6}), we can construct a surrogate function as
\begin{eqnarray*}
   Q(\bla|\blat) = \constant + \sum_{i=1}^n \sum_{k=0}^K \bt{ik} \log(\la_k) \pr \sum_{k=0}^K \left(\sum_{i=1}^n \bt{ik} \right) \log(\la_k),
\end{eqnarray*}
which satisfies $Q(\bla|\blat) \le \ell(\bla|\Y{obs})$ and $Q(\blat|\blat) = \ell(\blat|\Y{obs})$.

Finally, by maximizing the $Q(\bla|\blat)$ with respect to $\bla$, we obtain the MM iterations as
\begin{eqnarray} \label{zyffour2.8}
   \latI_k = \frac{\sum_{i=1}^n \bt{ik}} {\sum_{l=0}^K \sum_{i=1}^n \bt{il}} = \frac{\1_n^{\T} \ibbt{k}}{\1_n^{\T} \bBt \1_{K+1}}, \quad k \in \bbN_{0:K},
\end{eqnarray}
where $\bBt$ is an $n \times (K+1)$ matrix defined by
$$
   \bBt \teq (\bt{ik}) = (\ibbt{0}, \ibbt{1}, \dots, \ibbt{K}) \qwi \ibbt{k} = \threev{\bt{1k}}{\vdots}{\bt{nk}}
$$
being the $k$-column of $\bBt$, here $\bt{ik}$ is given by (\ref{zyffour2.7}). The MM iterations are repeated until
$$
   \bigg|\frac{\ell(\blatI|\Y{obs}) - \ell(\blat|\Y{obs})} {\ell(\blat|\Y{obs})}\bigg| \le \de_0.
$$
where $\de_0$ is a pre-specified small positive number.

\subsection{Asymptotic normality of MLEs and CIs of parameters}      % Section 2.3

Due to the existence of the constraint $\la_0 = 1 - \sum_{k=1}^K \la_k$ in $\bla = (\la_0, \la_1, \dots, \la_K)^{\T}$, we consider the parameter sub-vector $\bla_{-0} = (\la_1, \dots, \la_K)^{\T}$, where $\la_k \in (0, 1)$ (we do not consider the cases of parameter components with $\la_k = 0$ or $\la_k=1$, indicating that the $k$-th weight is zero or one). One of the key questions in large--sample statistical inferences is to verify the asymptotic normality of MLEs; i.e.,
\begin{eqnarray*}
   \sqrt{n} \left(\hbla_{-0, n} - \bla_{-0, \rm true} \right) \srto{D} N_K(\0, \bSi) \qas n \to \infty,
\end{eqnarray*}
where $\hbla_{-0, n}$ denotes the MLE of $\bla_{-0}$ for emphasizing the dependency of $\hbla_{-0}$ on the sample size $n$,  $\bla_{-0, \rm true}$ is the true value of $\bla_{-0}$ and $\bSi$ is the variance--covariance matrix of the asymptotic normal  distribution. The asymptotic normality depends on a series of assumptions known as regularization or Cram\'er--Rao conditions (\cite{LC1998} and \cite{Lehmann1999}), which are presented in Theorem \ref{thm2} of Appendix A.1. By verifying the Cram\'er--Rao conditions for Bern-Bino distribution (see Appendix A.2 for more details), we obtain the asymptotic normality of $\hbla_{-0, n}$ as
\begin{eqnarray} \label{zyffour2.9}
   \sqrt{n}\left(\hbla_{-0, n} - \bla_{-0, \rm true}\right) \srto{D}
   N_K \left(\0, \bJ^{-1}(\bla_{-0,\rm true}) \right) \qas n \to \infty,
\end{eqnarray}
where $\bJ(\bla_{-0,\rm true})$ is the Fisher information matrix. After obtaining (\ref{zyffour2.9}), by using the delta method, we obtain the asymptotic distribution of $\hla_0 \teq g(\hbla_{-0}) = 1 - \sum_{k=1}^K \hla_k$ as
$$
   \sqrt{n} (\hla_{0, n} - \la_{0,\rm true}) \srto{D}
   N \Big(0, [\na g(\bla_{-0, \rm true})]^{\T} \bJ^{-1}(\bla_{-0, \rm true}) \na g(\bla_{-0, \rm true}) \Big).
$$
In practice, we adopt the sandwich covariance matrix estimation to replace $\bJ^{-1}(\bla_{-0, \rm true})$, which can be computed as
$$
   \whbSi = \whbH^{-1} \whbS \whbH^{-1},
$$
where $\whbS$ is the outer product of scores and $\whbH$ is the observed information matrix
\begin{eqnarray*}
   \whbS &=& \frac{1}{n} \sum_{i=1}^n \ibs_i\ibs_i^{\T} \qwi \ibs_i = \na \log p(x_i|\bla_{-0}) \big|_{\bla_{-0} = \hbla_{-0,n}} \qand \\ [2mm]
   \whbH &=& -\na^2 \log \ell(\bla_{-0}|\Y{obs})\big|_{\bla_{-0} = \hat{\bla}_{-0,n}}.
\end{eqnarray*}
Finally, we construct the $100(1-\al)\%$ asymptotic CI of $\la_k$ as $\hla_{k,n} \pm z_{\al/2} \cdot \whse_k$ with standard error $\whse_k = \sqrt{\whbSi_{kk}}$ for $k \in \bbN_{1:K}$ and $\whse_0 = \sqrt{ [\na g(\hbla_{-0})]^{\T} \whbSi \na g(\hbla_{-0}) }$.

\subsection{Bootstrap confidence intervals of $\bla$}  % Section 2.4

The large--sample method to calculate CIs of parameters has two  limitations: (1) The asymptotic CI is not reliable for small to moderate sample sizes; and (2) even though for a large sample size, the range of CI may be beyond its bounds when the true value is close to the bounds. To overcome the two difficulties, this subsection presents the bootstrap method (\cite{Efron1992}) to construct the \textit{bootstrap confidence interval} (BCI, \cite{Hall1988}, \cite{DR1988}) of $\vth \teq h(\bla)$, where $h(\cdot)$ is an arbitrary function of the parameter vector $\bla$ in the $\BernBino(m, K, \bla)$ distribution.

Based on the observations $\Y{obs} = \{x_i\}_{i=1}^n$, we can first compute the MLEs $\hbla$ of $\bla$ through the MM algorithm (\ref{zyffour2.8}). Then, we generate a bootstrap sample $\{X_i^*= x_i^*\}_{i=1}^n \iid\BernBino(m, K, \hbla)$ and compute one bootstrap replication $\hvth^* = h(\hblas)$. Independently repeating this process $G$ times, we obtain $G$ bootstrap replications $\{\hvth^*(g)\}_{g=1}^G$. Hence, a $100(1-\al)$\% BCI of $\vth$ is
$[\hvth_{\rm L}^*, \  \hvth_{\rm U}^*]$, where $\hvth_{\rm L}^*$ and $\hvth_{\rm U}^*$ are the $(\al/2)G$-th and the $(1-\al/2)G$-th order statistics of $\{\hvth^*(g)\}_{g=1}^G$.

\subsection{Regression model}      % Section 2.5

Assume that $\{X_i\}_{i=1}^n \ind \BernBino(m, K, \bla_i)$ and the observed data be $\Y{obs} = \{x_i, \ibzi\}_{i=1}^n$, where $\ibzi = (z_{i1}, \ldots, z_{ip})^\top$ is the covariate vector of the $i$-th subject. To reduce the number of coefficients and ensure the identifiability, we consider the following regression model
\begin{eqnarray}
    && \la_{ik} = \frac{\exp(\eta_{ik})}{\sum_{l=0}^K \exp(\eta_{il})}, \quad i \in \bbN_{1:n}, \qwh \label{zyf.BSB.2.12} \\ [2mm]
    && \eta_{ik} = \phi_k + \ga_k \cdot \ibzi^\top \bbe, \quad k \in \bbN_{0:K} \qwi \phi_0 = 0, \; \ga_0 = 1. \label{zyf.BSB.2.13}
\end{eqnarray}
The log-likelihood function of $\bth = (\bphi^{\T}, \bga^{\T}, \bbe^{\T})^{\T}$ is given by
$$
    \ell(\bth|\Y{obs}) = \sum_{i=1}^n \log\binom{m}{x_i} + \sum_{i=1}^n \log \left( \sum_{k=0}^k a_{ik} \la_{ik} \right).
$$
By calculating the score vector $\ibs(\bth)$ and Hessian matrix $\bH(\bth)$, we adopt the Newton--Raphson algorithm to compute the MLE $\hbth$ of parameter $\bth$ as
$$
    \bthtt = \btht - \bH^{-1} (\btht) \ibs(\btht).
$$ 
%!TEX root = BSB.tex

\section{$\!\!\!\!\!\!\!$. Bayesian inferences} % Section 3

Let $\{X_i\}_{i=1}^n \iid \BernBino(m, K, \bla)$ and the observed data $\Y{obs} = \{x_i\}_{i=1}^n$. We perform Bayesian inferences (\cite{BT1992}) on parameters $\bla = (\la_0, \la_1, \dots, \la_K)^{\T} \in \bbT_{K+1}$; i.e., we will calculate the posterior mode via an MM algorithm and generate posterior samples via a DA algorithm. Finally, the posterior mean and Bayesian credible interval can be obtained based on posterior samples.

Consider the prior $\bla \sim \Dirichlet_{K+1}(\bal)$ with joint pdf
\begin{eqnarray}    \label{eqn2.4}
   \Dirichlet_{K+1}(\bla| \bal) = \frac{\Ga(\sum_{k=0}^K \al_k)} {\prod_{k=0}^K \Ga(\al_k)} \prod_{k=0}^K \la_k^{\al_k-1}, \quad \bla \in \bbT_{K+1},
\end{eqnarray}
where $\bal = (\al_0,\al_1,\ldots,\al_K)^{\T}$ is a known vector. Then, we obtain the posterior distribution
\begin{eqnarray}    \label{eqn2.5}
   p(\bla|\Y{obs}) \pr \cL(\bla|\Y{obs}) \cdot \Dirichlet_{K+1}(\bla| \bal),
\end{eqnarray}
where $\cL(\bla|\Y{obs})$ is the likelihood function of $\bla$. Find the mode of posterior distribution is equivalent to optimizing  $\log p(\bla|\Y{obs})$ to obtain the optimized value $\tbla$.

\subsection{The MM algorithm to calculate the posterior mode of $\bla$}      % Section 3.1

From (\ref{eqn2.5}) with the prior (\ref{eqn2.4}), we easily obtain
\begin{eqnarray*}
   \log p(\bla|\Y{obs}) = \constant + \sum_{i=1}^n \log \left(\sum_{k=0}^K a_{ik}\la_k \right) + \sum_{k=0}^K (\al_k-1)\log (\la_k).
\end{eqnarray*}
Similar to \S2.2, we employ Jensen's inequality again to construct the surrogate function
$$
   Q(\bla|\blat) = \constant + \sum_{k=0}^K \left(\sum_{i=1}^n \bt{ik} + \al_k-1 \right) \log (\la_k),
$$
where $\bt{ik}$ is defined by (\ref{zyffour2.7}). Similar to (\ref{zyffour2.8}), we obtain the MM iterations as
\begin{eqnarray*}
   \latI_k = \frac{\sum_{i=1}^n \bt{ik}+\al_k-1}{\sum_{l=0}^K \Big(\sum_{i=1}^n \bt{il}+\al_l-1 \Big)}, \quad k \in \bbN_{0:K}.
\end{eqnarray*}

\subsection{The DA algorithm to generate posterior samples}     % Section 3.2

From (\ref{zyffour2.3}), we know that $\ibZ =(Z_0, Z_1, \dots, Z_K)^{\T} \sim \Multinomial(1,\bla)$ is equivalent to a finite discrete \trv $W$ defined on $\bbN_{0:K}$ with corresponding probabilities $\bla$, where $W = k$ means $Z_k  = 1$ is sampled from the $k$-th category. Therefore, we introduce the missing data $\Y{mis} = \{w_i\}_{i=1}^n$, where the latent variable $W_i \iid \FDiscrete(\bbN_{0:K}, \bla)$ with $p(W_i=k) = \la_k$ for $k \in \bbN_{0:K}$. Therefore, the complete data can be expressed as $\Y{com} = \{\Y{obs},\Y{mis}\}$.

In order to make a full Bayesian inference on the parameter vector $\bla$, we could employ the DA algorithm of Tanner \& Wong (1987), where the \textsf{I-step} is to generate the missing data $\{W_i=w_i\}_{i=1}^n$ from the conditional predictive distribution
\begin{eqnarray*}
    p(W_i = k | x_i, \bla) &\pr & p(x_i | W_i = k, \bla) \cdot p(W_i = k | \bla) \\ [2mm]
    &=& \la_k \cdot \binom{m}{x_i} \frac{B(x_i + k + 1, m - x_i + K - k + 1)}{B(k + 1, K - k + 1)} \\ [2mm]
    &=& \la_k \cdot \BetaBino(x_i | m, k + 1, K - k + 1) \teq q_{ik};
\end{eqnarray*}
i.e., $W_i | (x_i, \bla) \sim \FDiscrete(\bbN_{0:K}, \ibp_i)$, where
$\ibp_i \teq (p_{i0}, p_{i1}, \dots, p_{iK})^{\T}$ with $p_{ik} \pr q_{ik}$. The \textsf{P-step} is draw $\bla$ from the complete--data posterior distribution
\begin{eqnarray}
    p(\bla | \Y{com}) = p(\bla | \ibW) = \pi(\bla | \bal) \cdot p(\ibW | \bla) \pr \prod_{k=0}^K \la_k^{\al_k + N_k - 1} \sim \Dirichlet_{K+1}(\bal + \ibN), \nonumber
\end{eqnarray}
where $\ibW = (W_1, \ldots, W_n)^\top$ and $\ibN = (N_0, N_1, \ldots,N_K)^{\T}$ with $N_k = \sum_{i=1}^n I(W_i = k)$. Thus, we summarize the DA algorithm for calculating the posterior mean and Bayesian CIs of $\la_k$s as follows:

\vkL
\IncMargin{0.1em}  % 使得行号不向外突出
\begin{algorithm}[H]  \label{alg1}
    \SetAlgoNoLine % 不要算法中的竖线
    \SetKwInOut{Input}{\textsf{Input}}\SetKwInOut{Output}{\textsf{Output}} % 替换关键词

    \Input{The initial parameter vector $\bla^{(0)}$, latent variable $\ibW^{(0)}$, the observed data $\Y{obs}=\{x_i\}_{i=1}^n$ and the sampling number $G$.}
    \Output{Posterior mean $\Bla_k$ and Bayesian CI $[\la_k^{\rm L},\;\la_k^{\rm U}]$ for $k \in \bbN_{0:K}$.}
    \BlankLine

    Set initial vectors $\bla^{(0)}$ and $\ibW^{(0)}$; \\
    \For{ \(g=0\) {\rm \bf to} \(G\)} {
        \textsf{I-step}: Draw latent variables $\ibW^{(g+1)}$ from $W_i|\bla^{(g)} \sim \FDiscrete(\bbN_{0:K}, \ibp_i^{(g)})$, $i \in \bbN_{1:n}$, where $\ibp_i^{(g)} = (p_{i0}^{(g)}, p_{i1}^{(g)}, \dots, p_{iK}^{(g)})^{\T}$  with $p_{ik}^{(g)} \pr q_{ik}^{(g)}$.

        \textsf{P-step}: Draw parameter vector $\bla^{(g+1)}$ from $\bla|\ibW^{(g+1)} \sim \Dirichlet_{K+1}(\bal+ \ibN^{(g+1)})$, where $\ibN^{(g)} = (N_0^{(g)}, N_1^{(g)}, \ldots, N_K^{(g)})^{\T}$ with $N_k^{(g)} = \sum_{i=1}^n I(W_i^{(g)} = k)$.
    }
    \textbf{Collection of posterior samples:} Abandon the results of the previous $B$ iterations and save the remaining samples $\{\bla^{(g)}\}_{g=B+1}^G$.

    \textbf{Compute the posterior mean:} $\Bla_k \teq [1/(G-B)]\sum_{g=B+1}^G \la_k^{(g)}$, $k\in \bbN_{0:K}$.

    \textbf{Compute the Bayesian CI:} $[\la_k^{\rm L},\; \la_k^{\rm U}] = [\la_k^{(\al/2)},\;\la_k^{(1-\al/2)}]$, $k\in \bbN_{0:K}$, where $\la_k^{(q)}$ is the $q$-quantile of the posterior sample.
    \caption{\ Posterior mean and Bayesian CI via the DA algorithm}
\end{algorithm}
\DecMargin{0.1em} 
%!TEX root = BSB.tex

\section{$\!\!\!\!\!\!\!$. Simulation studies}    % Section 4

\subsection{Accuracy of MLEs of parameters and reliability of CIs}     % Section 4.1

In \S 2, we proposed the $\BernBino(m, K, \bla)$ distribution and calculated the MLEs of $\bla$ via an MM algorithm. In addition, the CIs of parameters can be obtained by the sandwich method for large sample sizes or by the bootstrap method for the small to moderate sample sizes. To assess the performance of the proposed algorithm and methods, we conduct some numerical experiments by employing the \textit{mean square error} (\textsf{MSE}) as the criterion to evaluate the accuracy of the MLE of each parameter and apply the \textit{coverage rate} (\textsf{CR}) and \textit{interval width} (\textsf{I-Wid}) to confirm the reliability of CIs. Consider the sample sizes to be $n=200(200)$1,000 and the number of sampling replications to be $R = $ 10,000 in our simulations. We conduct some numerical experiments for the following two cases:
\begin{namelist}{0123456789}
\item[\hspace*{-0.1cm} Case ($\mbox{A}_1$):] Independently generating $\{X_i^{(r)}\}_{i=1}^n \iid \BernBino(m, K, \bla)$ with the true values being set as $m = 8$, $K=2$, $\bla = (0.2, 0.5, 0.3)^{\T}$;

\item[\hspace*{-0.1cm} Case ($\mbox{A}_2$):] Independently generating $\{X_i^{(r)}\}_{i=1}^n \iid \BernBino(m, K, \bla)$ with the true values being set as $m = 15$, $K=3$, $\bla = (0.15, 0.3, 0.35, 0.2)^{\T}$;
\end{namelist}

Given a combination of $\{n, r, m, K, \bla\}$ with $r \in \bbN_{1:R}$, for each case, let $\Y{obs}^{(r)} = \{x_i^{(r)}\}_{i=1}^n$ denote the $r$-th generated observations, where $x_i^{(r)}$ is a realization of $X_i^{(r)}$. We can compute the MLEs of the parameter vector $\bla$ via the MM algorithm (\ref{zyffour2.8}), denoted by $\hbla^{(r)}$. Next, we calculate the \textit{average MLE} (\textsf{A-MLE}), \textsf{MSE} based on the $R$ repetitions. Finally, we calculate the \textsf{CR} and \textsf{I-Wid} of CIs, which are obtained by the sandwich method. All these results are presented in Table \ref{zyffour.tab1}.

\newpage
\vkU
\baselineskip 0.10in
\begin{Table} \label{zyffour.tab1} {\rm \hspace{-0.2cm}{\bf .}
Comparisons of \textsf{A-MLE}, \textsf{MSE}, \textsf{CR} and \textsf{I-Wid} based on $R=$ 10,000 sampling replications in Bern-Bino distributions for Cases ($\mbox{A}_1$)--($\mbox{A}_2$) with various sample sizes
\vspace{-0.2cm}
\begin{center}
\renewcommand{\arraystretch}{1.2} \tabcolsep 0.15in \doublerulesep 0.2pt
\begin{tabular}{|l|c|l|r|r|r|r|r|}  \hline\hline
  \MR{2}{*}{Case} & \MR{2}{*}{Parameter} & \MR{2}{*}{Criterion} & \MC{5}{c|}{Sample size} \\ \cline{4-8}
  \MR{2}{*}{} & \MR{2}{*}{} & \MR{2}{*}{} & \MC{1}{c|}{200} & \MC{1}{c|}{400} & \MC{1}{c|}{600} & \MC{1}{c|}{800} & \MC{1}{c|}{1000} \\ \hline
  \MR{12}{*}{($\mbox{A}_1$)} & \MR{4}{*}{$\la_0$} & \textsf{\textsf{A-MLE}} & 0.2021 & 0.2011 & 0.2005 & 0.2007 & 0.2011 \\
  \MR{12}{*}{}               & \MR{4}{*}{}            & \textsf{MSE}   & 0.0065 & 0.0032 & 0.0021 & 0.0016 & 0.0013 \\
  \MR{12}{*}{}               & \MR{4}{*}{}            & \textsf{CR}    & 0.9445 & 0.9525 & 0.9497 & 0.9475 & 0.9462 \\
  \MR{12}{*}{}               & \MR{4}{*}{}            & \textsf{I-Wid} & 0.3156 & 0.2237 & 0.1827 & 0.1583 & 0.1416 \\ \cline{2-8}
  \MR{12}{*}{}               & \MR{4}{*}{$\la_1$} & \textsf{A-MLE} & 0.4966 & 0.4980 & 0.4991 & 0.4985 & 0.4980 \\
  \MR{12}{*}{}               & \MR{4}{*}{}            & \textsf{MSE}   & 0.0203 & 0.0102 & 0.0067 & 0.0051 & 0.0042 \\
  \MR{12}{*}{}               & \MR{4}{*}{}            & \textsf{CR}    & 0.9475 & 0.9506 & 0.9503 & 0.9492 & 0.9458 \\
  \MR{12}{*}{}               & \MR{4}{*}{}            & \textsf{I-Wid} & 0.5606 & 0.3973 & 0.3245 & 0.2811 & 0.2515 \\ \cline{2-8}
  \MR{12}{*}{}               & \MR{4}{*}{$\la_2$} & \textsf{A-MLE} & 0.3013 & 0.3009 & 0.3004 & 0.3009 & 0.3008 \\
  \MR{12}{*}{}               & \MR{4}{*}{}            & \textsf{MSE}   & 0.0075 & 0.0038 & 0.0025 & 0.0019 & 0.0016 \\
  \MR{12}{*}{}               & \MR{4}{*}{}            & \textsf{CR}    & 0.9476 & 0.9532 & 0.9477 & 0.9502 & 0.9465 \\
  \MR{12}{*}{}               & \MR{4}{*}{}            & \textsf{I-Wid} & 0.3403 & 0.2412 & 0.1970 & 0.1707 & 0.1527 \\ \hline
  \MR{16}{*}{($\mbox{A}_2$)} & \MR{4}{*}{$\la_0$} & \textsf{A-MLE} & 0.1493 & 0.1488 & 0.1498 & 0.1492 & 0.1496 \\
  \MR{16}{*}{}               & \MR{4}{*}{}            & \textsf{MSE}   & 0.0054 & 0.0029 & 0.0020 & 0.0015 & 0.0012 \\
  \MR{16}{*}{}               & \MR{4}{*}{}            & \textsf{CR}    & 0.9168 & 0.9408 & 0.9497 & 0.9524 & 0.9499 \\
  \MR{16}{*}{}               & \MR{4}{*}{}            & \textsf{I-Wid} & 0.3492 & 0.2287 & 0.1805 & 0.1570 & 0.1396 \\ \cline{2-8}
  \MR{16}{*}{}               & \MR{4}{*}{$\la_1$} & \textsf{A-MLE} & 0.3024 & 0.3025 & 0.3006 & 0.3024 & 0.3016 \\
  \MR{16}{*}{}               & \MR{4}{*}{}            & \textsf{MSE}   & 0.0315 & 0.0186 & 0.0127 & 0.0093 & 0.0077 \\
  \MR{16}{*}{}               & \MR{4}{*}{}            & \textsf{CR}    & 0.8789 & 0.9444 & 0.9578 & 0.9586 & 0.9520 \\
  \MR{16}{*}{}               & \MR{4}{*}{}            & \textsf{I-Wid} & 0.8380 & 0.5784 & 0.4552 & 0.3970 & 0.3525 \\ \cline{2-8}
  \MR{16}{*}{}               & \MR{4}{*}{$\la_2$} & \textsf{A-MLE} & 0.3490 & 0.3480 & 0.3504 & 0.3474 & 0.3487 \\
  \MR{16}{*}{}               & \MR{4}{*}{}            & \textsf{MSE}   & 0.0338 & 0.0199 & 0.0135 & 0.0097 & 0.0081 \\
  \MR{16}{*}{}               & \MR{4}{*}{}            & \textsf{CR}    & 0.8833 & 0.9479 & 0.9522 & 0.9581 & 0.9534 \\
  \MR{16}{*}{}               & \MR{4}{*}{}            & \textsf{I-Wid} & 0.8181 & 0.5934 & 0.4674 & 0.4075 & 0.3620 \\ \cline{2-8}
  \MR{16}{*}{}               & \MR{4}{*}{$\la_3$} & \textsf{A-MLE} & 0.1993 & 0.2007 & 0.1993 & 0.2010 & 0.2001 \\
  \MR{16}{*}{}               & \MR{4}{*}{}            & \textsf{MSE}   & 0.0066 & 0.0036 & 0.0024 & 0.0018 & 0.0015 \\
  \MR{16}{*}{}               & \MR{4}{*}{}            & \textsf{CR}    & 0.9219 & 0.9448 & 0.9486 & 0.9555 & 0.9486 \\
  \MR{16}{*}{}               & \MR{4}{*}{}            & \textsf{I-Wid} & 0.3664 & 0.2471 & 0.1957 & 0.1704 & 0.1517 \\ \hline \hline
\end{tabular}
\end{center} }
\end{Table}
\vspace{-0.2cm}
\noi  {\small  The convergence precision of \textsf{A-MLE} is $10^{-8}$.}
\baselineskip 0.30in

\vkL From the results in Table \ref{zyffour.tab1}, we can found that the proposed MM algorithm is very accurate in estimating the parameters. As the sample size increases, the \textsf{MSE} and \textsf{I-Wid} gradually decreases, which makes the estimation more robust. However, in the case of small samples, the asymptotic normality of the parameters is poor, so the problem of insufficient interval coverage rate may occur when the number of parameters is large (like Case ($\mbox{A}_2$) with sample size $n=200$). In such case, we can select the bootstrap method as an alternative to obtain the accurate CI.

\subsection{Comparison of Bernstein prior generalization}     % Section 4.2

In order to systematically evaluate the performance of the our new distribution constructed based on Bernstein prior in complex data scenarios, a series of numerical simulation experiments are designed and carried out in this subsection. The purpose of this study is to compare and analyze the advantages and limitations of Bern-Bino distribution and existing mainstream distribution in terms of fitting accuracy and robustness by simulating complex data from different real prior distributions. Specifically, we will generate the success probability parameter $p$ of binomial distribution from several common distributions with different morphological characteristics, and then generate the corresponding binomial discrete observation samples, so as to test the ability of the new model in capturing the potential a prior structural complexity and mitigating the deviation of model setting.

To comprehensively and objectively quantify the fitting effect of each model, we use multiple evaluation criteria for comprehensive comparison: including the log-likelihood function value of the model, which is used to directly measure the fitting degree of the model to the data. \textit{Akaike information criterion} (AIC, \cite{Akaike1974}), which introduces penalty term on the basis of likelihood to balance model complexity and goodness of fit and avoid over fitting. And $\chi^2$ goodness of fit test, the absolute fitting effect is evaluated by comparing the observation frequency with the model prediction frequency, where the corresponding $p$-value is provided as the judgment basis.

In the experimental design, the sample size, the number of trials in binomial distribution, the number of sampling replications are respectively set to be $n = 500$, $m = 30$ and $R=$ 1,000. We selected four representative distributions as the real prior of generating binomial data success probability $p$ to simulate the possible diversity of prior information in the real world: beta distribution, logit-normal (Logit-N) distribution, and two complex mixed models called Logistic--Stdlogistic (LS) distribution and Kumaraswany (KM) distribution. For a given $\{n, \; r, \; \bth\}$ with $r = 1, \ldots, R$ and $\bth = (\al_1, \be_1, \mu_2, \sigma_2, \mu_3, \sigma_3, \al_4, \be_4)^{\T}$, we independently generate the prior $P_i^{(r)}=p_i^{(r)}$ from the mixed distribution:
$$
    \om_1 \cdot \mBeta(\al_1,\be_1) + \om_2 \cdot \mbox{Logit-N}(\mu_2,\sigma_2) + \om_3 \cdot \mbox{LS}(\mu_3,\sigma_3) + \om_4 \cdot \mbox{KM}(\al_4,\be_4),
$$
where $\om_s \ge 0$ for $s = 1, \dots, 4$ with $\sum_{s=1}^4 \om_s = 1$. We consider the following situations for selecting $\bom = (\om_1, \dots, \om_4)^{\T}$ and parameters in four prior distribution:
\begin{namelist}{0123456789}

\item[\hspace*{-0.1cm} Case ($\mbox{B}_1$):] Uniform four–component mixed distribution: $\bom = 0.25 \times \1_4$ with parameters $\bth = (1, 3, -1, 0.5, 2, 5, 1, 2)^{\T}$;

\item[\hspace*{-0.1cm} Case ($\mbox{B}_2$):] Dominant one–component mixed distribution: One $\om_s = 0.4$, while others $\om_{s'} = 0.2$ for $s' \ne s$ with parameters $\bth = (1, 3, -1, 0.5, 2, 5, 1, 2)^{\T}$;

\item[\hspace*{-0.1cm} Case ($\mbox{B}_3$):] A complex mixed distribution with $\bom = 0.25 \times \1_4$ and parameter vector $\bth = (1, 0.5, 0, 0.25, 2, 5, 1, 5)^\top$.
\end{namelist}
By sampling from the mixed distribution, based on the generated prior $\{p_1^{(r)}, \dots, p_n^{(r)}\}$, we further generate the corresponding binomial distribution sample $\{x_1^{(r)}, \dots, x_n^{(r)}\}$. Then we fit the corresponding binomial mixture distribution and obtain their log-likelihoods, AICs, $p$-values for $\chi^2$ goodness of fit test, which is proposed by Karl Pearson (1900) to evaluate whether the deviation between the observation frequency ($O_j$) and the expected frequency ($E_j$) under the theoretical distribution is caused by random sampling error. The test statistics are the sum of squares of standardized residuals:
$$
    \chi_0^2 = \sum_{j = 0}^m \frac{(O_j - E_j)^2}{E_j},
$$
and the corresponding $p$-value $= \Pr(\chi^2 > \chi_0^2 \; | \;\nu)$ with degree of freedom $\nu$.

We obtain the corresponding five ranks $\{\mbox{Rank}_{1,A}^{(r)}, \ldots, \mbox{Rank}_{5,A}^{(r)}\}$, where $\mbox{Rank}_{s,A}^{(r)}$ represents the rank of $\AIC_s^{(r)}$ in $\{\AIC_1^{(r)},\dots,\AIC_5^{(r)}\}$. And $\{\mbox{Rank}_{1,p}^{(r)}, \ldots, \mbox{Rank}_{5,p}^{(r)}\}$ represents the rank of $\{\mbox{\textit{p}-value}_1^{(r)}, \dots, \mbox{\textit{p}-value}_5^{(r)}\}$. Finally, the average log-likelihoods (denoted by $\textsf{Log-likelihood}$), AICs (denoted by \textsf{AIC}), $p$-value (\textsf{p-value}) and two ranks ($\textsf{Rank}_A$, $\textsf{Rank}_p$) are calculated by
\begin{eqnarray}
    &&\textsf{Log-likelihood}_s = \frac{1}{R}\sum_{r=1}^{R} \textsf{Log-likelihood}_s^{(r)}, \quad \textsf{AIC}_s = \frac{1}{R}\sum_{r=1}^{R} \AIC_s^{(r)} \non \\ [2mm]
    &&\textsf{p-value}_s = \frac{1}{R}\sum_{r=1}^{R} \textsf{p-vlaue}_s^{(r)} \qand \textsf{Rank}_{s,t} = \frac{1}{R}\sum_{r=1}^{R}\mbox{Rank}_{s,t}^{(r)},\quad s=1,\dots,5, \ t = A,p. \non
\end{eqnarray}
Table \ref{tab2} presents the \textsf{Log-likelihood}s, \textsf{AIC}s, $\textsf{Rank}_A$s, \textsf{p-value}s and $\textsf{Rank}_p$s of five fitted distributions. From the table, we can found that, in Cases ($\mbox{B}_1$) and ($\mbox{B}_2$), Bernstein prior has excellent fitting effect on the simulated data. The log-likelihood value of Bernstein prior fitting is the largest, which means Bern-Bino distribution has the strongest ability to interpret the data. And the AIC value is the smallest among all models, indicating that it achieves a good balance between fitting accuracy and complexity. At the same time, the $p$-value of chi square test of Bern-Bino distribution is also the largest, therefore, there is no significant difference between the observed data and the distribution we fit. These results show that in the common mixed prior scenario, the performance of the Bern-Bino distribution is not inferior to that of traditional models, but rather it has more advantages in fitting robustness and generalization ability. And in the complex mixed situation of Case ($\mbox{B}_3$), severe fitting failures occurred in various traditional simple distributions, the $p$-values of their $\chi^2$ tests are almost zero. This indicates that these distributions are no longer effectively capable the complex prior heterogeneous structures. In contrast, the Bern-Bino distribution still maintains excellent fitting performance. Its log-likelihood value is significantly higher, the AIC is significantly lower, and the p-value reaches 0.4863, which is much higher than that of other distributions, which fully demonstrates that Bern-Bino distribution can still stably and accurately depict the true form under complex structure. In conclusion, we think Bern-Bino distribution perform excellent with outstanding flexibility and robustness.

\newpage
\vkU
\baselineskip 0.10in
\begin{Table} \label{tab2} {\rm \hspace{-0.2cm}{\bf .}
Comparison of \textsf{Log-likelihood}, \textsf{AIC}, $\textsf{Rank}_A$, \textsf{p-value} and $\textsf{Rank}_p$ based on $R = $ 1,000 sampling replications for five distributions in Cases ($\mbox{B}_1$)--($\mbox{B}_3$).
\vspace{-0.2cm}
\begin{center}
\renewcommand{\arraystretch}{1.2} \tabcolsep 0.1in \doublerulesep 0.2pt
\begin{tabular}{|l|l|c|c|c|c|c|}  \hline\hline
    Case              & \MC{1}{l|}{Prior} & \textsf{Log-likelihood} & \textsf{AIC}      & $\textsf{Rank}_A$ & \textsf{p-value} & $\textsf{Rank}_p$ \\ \hline
    \MR{5}{*}{($\mbox{B}_1$)} & Beta              & $-$1559.059             & 3122.119          & 3.2401            & 0.1199           & 3.0947            \\
    \MR{5}{*}{}               & Logit-N           & $-$1565.753             & 3135.507          & 1.0322            & 0.0318           & 1.3030            \\
    \MR{5}{*}{}               & LS                & $-$1561.722             & 3127.443          & 2.3897            & 0.0918           & 2.5592            \\
    \MR{5}{*}{}               & KM                & $-$1558.132             & 3120.265          & 4.2008            & 0.1248           & 3.0507            \\
    \MR{5}{*}{}               & Bernstein         & $-$1551.174             & 3118.432          & 4.1373            & 0.5841           & 4.9924            \\ \hline
    \MR{5}{*}{\makecell{($\mbox{B}_2$) with $\bom = $ \\ $(0.4,0.2,0.2,0.2)^{\T}$}} & Beta      & $-$1558.663 & 3121.326 & 3.4688 & 0.1768 & 3.2202 \\
    \MR{5}{*}{}                                                                     & Logit-N   & $-$1566.051 & 3136.102 & 1.1113 & 0.0406 & 1.2893 \\
    \MR{5}{*}{}                                                                     & LS        & $-$1563.694 & 3131.387 & 2.0374 & 0.0837 & 2.2088 \\
    \MR{5}{*}{}                                                                     & KM        & $-$1557.889 & 3119.777 & 4.4555 & 0.1852 & 3.2945 \\
    \MR{5}{*}{}                                                                     & Bernstein & $-$1552.819 & 3119.563 & 3.9271 & 0.5620 & 4.9872 \\ \hdashline
    \MR{5}{*}{\makecell{($\mbox{B}_2$) with $\bom = $ \\ $(0.2,0.4,0.2,0.2)^{\T}$}} & Beta      & $-$1534.893 & 3073.787 & 3.2841 & 0.0923 & 2.8423 \\
    \MR{5}{*}{}                                                                     & Logit-N   & $-$1539.508 & 3083.015 & 1.1463 & 0.0433 & 1.7372 \\
    \MR{5}{*}{}                                                                     & LS        & $-$1534.770 & 3073.540 & 3.4252 & 0.1597 & 3.2098 \\
    \MR{5}{*}{}                                                                     & KM        & $-$1534.540 & 3073.080 & 3.6070 & 0.0814 & 2.2315 \\
    \MR{5}{*}{}                                                                     & Bernstein & $-$1527.199 & 3071.843 & 3.5374 & 0.5902 & 4.9792 \\ \hdashline
    \MR{5}{*}{\makecell{($\mbox{B}_2$) with $\bom = $ \\ $(0.2,0.2,0.4,0.2)^{\T}$}} & Beta      & $-$1553.071 & 3110.141 & 2.9413 & 0.0362 & 3.0791 \\
    \MR{5}{*}{}                                                                     & Logit-N   & $-$1560.936 & 3125.873 & 1.0019 & 0.0074 & 1.6117 \\
    \MR{5}{*}{}                                                                     & LS        & $-$1555.392 & 3114.784 & 2.3627 & 0.0208 & 2.0559 \\
    \MR{5}{*}{}                                                                     & KM        & $-$1550.760 & 3105.519 & 4.1136 & 0.0473 & 3.2571 \\
    \MR{5}{*}{}                                                                     & Bernstein & $-$1540.049 & 3099.329 & 4.5805 & 0.5029 & 4.9962 \\ \hdashline
    \MR{5}{*}{\makecell{($\mbox{B}_2$) with $\bom = $ \\ $(0.2,0.2,0.2,0.4)^{\T}$}} & Beta      & $-$1580.142 & 3164.284 & 3.4138 & 0.1786 & 3.1506 \\
    \MR{5}{*}{}                                                                     & Logit-N   & $-$1586.090 & 3176.180 & 1.0393 & 0.0543 & 1.1809 \\
    \MR{5}{*}{}                                                                     & LS        & $-$1582.358 & 3168.717 & 2.5563 & 0.1518 & 2.7334 \\
    \MR{5}{*}{}                                                                     & KM        & $-$1579.774 & 3163.549 & 4.0417 & 0.1777 & 2.9455 \\
    \MR{5}{*}{}                                                                     & Bernstein & $-$1574.369 & 3162.118 & 3.9489 & 0.5741 & 4.9896 \\ \hdashline
    \MR{5}{*}{($\mbox{B}_3$)}   & Beta            & $-$1702.888    & 3409.775 & 2.8097          & 0.0005    & 2.5005 \\
    \MR{5}{*}{}                 & Logit-N         & $-$1714.498    & 3432.997 & 1.0000          & 0.0005    & 2.5005 \\
    \MR{5}{*}{}                 & LS              & $-$1697.513    & 3399.026 & 4.0000          & 0.0005    & 2.5024 \\
    \MR{5}{*}{}                 & KM              & $-$1702.944    & 3409.888 & 2.1903          & 0.0005    & 2.5005 \\
    \MR{5}{*}{}                 & Bernstein       & $-$1653.137    & 3330.307 & 5.0000          & 0.4863    & 4.9962 \\ \hline
\end{tabular}
\end{center} }
\end{Table}
\baselineskip 0.30in

In addition, we also generate sample and fit distributions under large samples to explore the performance of Bernstein--binomial distribution for Cases ($\mbox{B}_1$) and ($\mbox{B}_3$), we set the sample size $n = $ 5,000 to generate mixed sample and use the above five distributions to perform the fitting and comparison.
\begin{figure}[H]
    \centering
    \subfigure[$\;$ Case ($\mbox{B}_1$)]
    {
        \centering
        \includegraphics[width=0.479\linewidth]{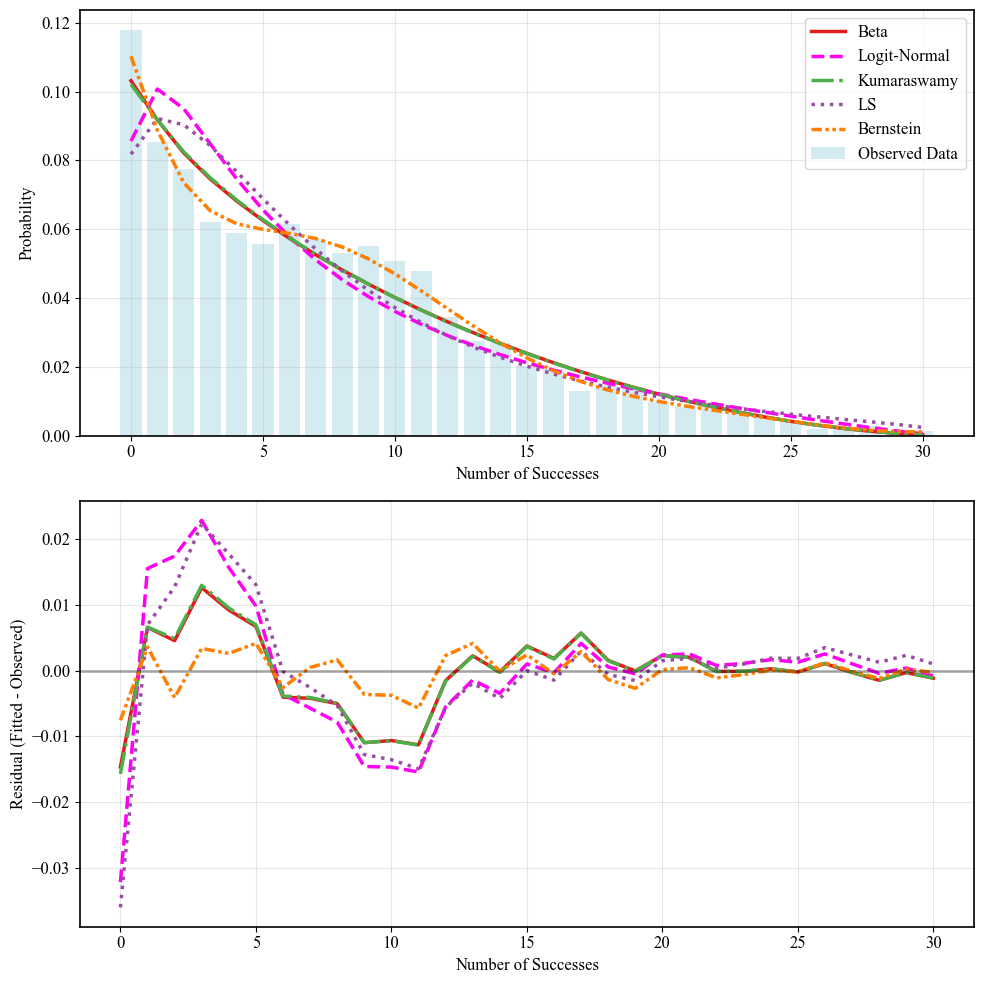}
    }
    \subfigure[$\;$ Case ($\mbox{B}_3$)]
    {
        \centering
        \includegraphics[width=0.479\linewidth]{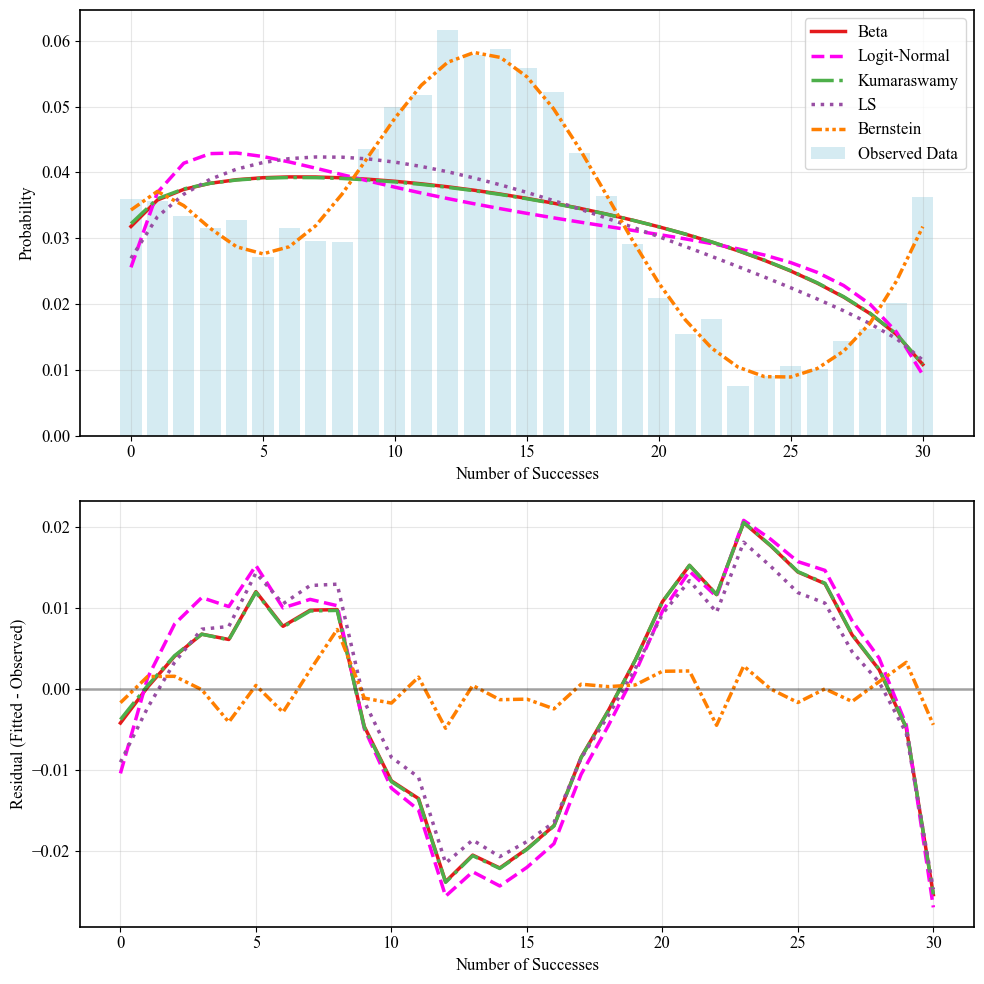}
    }
    \caption{$\;$ Fitting effect of five distributions and the corresponding residual plots.}
\end{figure}
Figure 2 shows the histogram of the generated samples together with the fitting curve of each distribution, and the residuals of the model fitting, which visually shows the proximity of each model to the empirical distribution. Sub-figure (a) and (b) in Figure 3 uses the Log-likelihood value, AIC, BIC, $p$-value of $\chi^2$ goodness test, \textit{Mean absolute error} (MAE) as evaluation criteria to evaluate the fitting effect of five models, and the comprehensive performance of each distribution are clearly displayed by radar chart. We can see that Bern-Bino has the largest area in the radar chart, which performs best. Sub-figure (c) and (d) represents the Bern-Bino distribution gradually approximates the real distribution with the increase of $K$, showing the flexibility and convergence.
\begin{figure}[H]
    \centering
    \subfigure[$\;$ Radar chart of Case ($\mbox{B}_1$)]
    {
        \centering
        \includegraphics[width=0.479\linewidth]{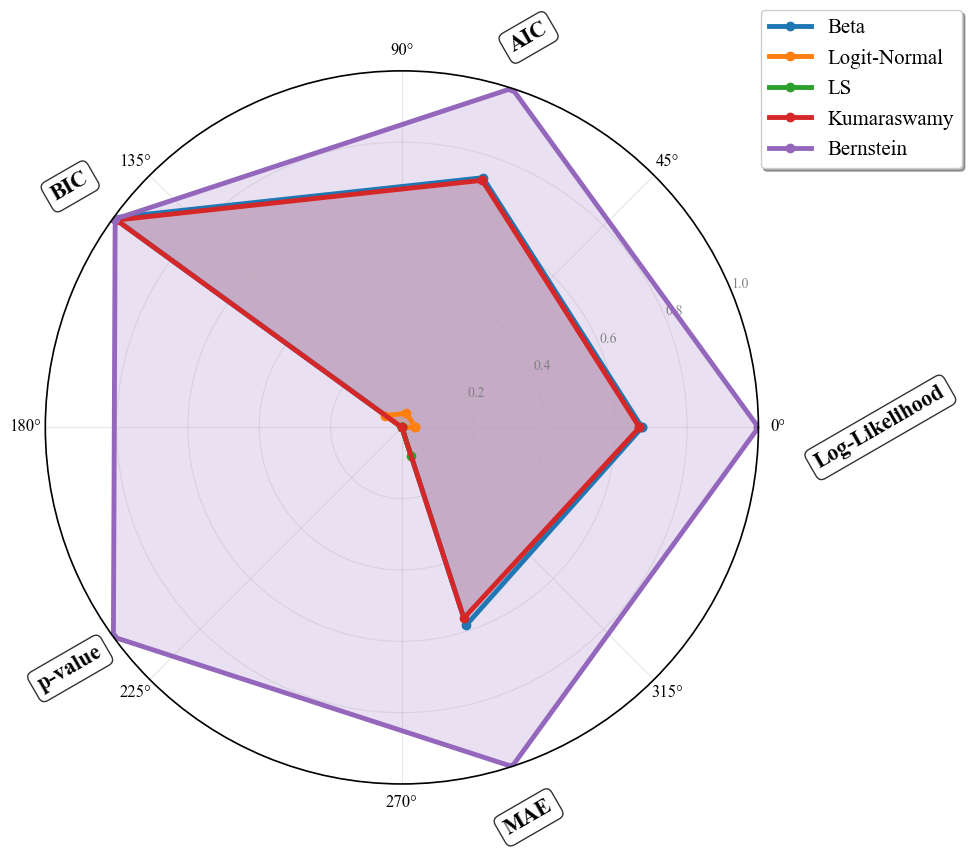}
    }
    \subfigure[$\;$ Radar chart of Case ($\mbox{B}_3$)]
    {
        \centering
        \includegraphics[width=0.479\linewidth]{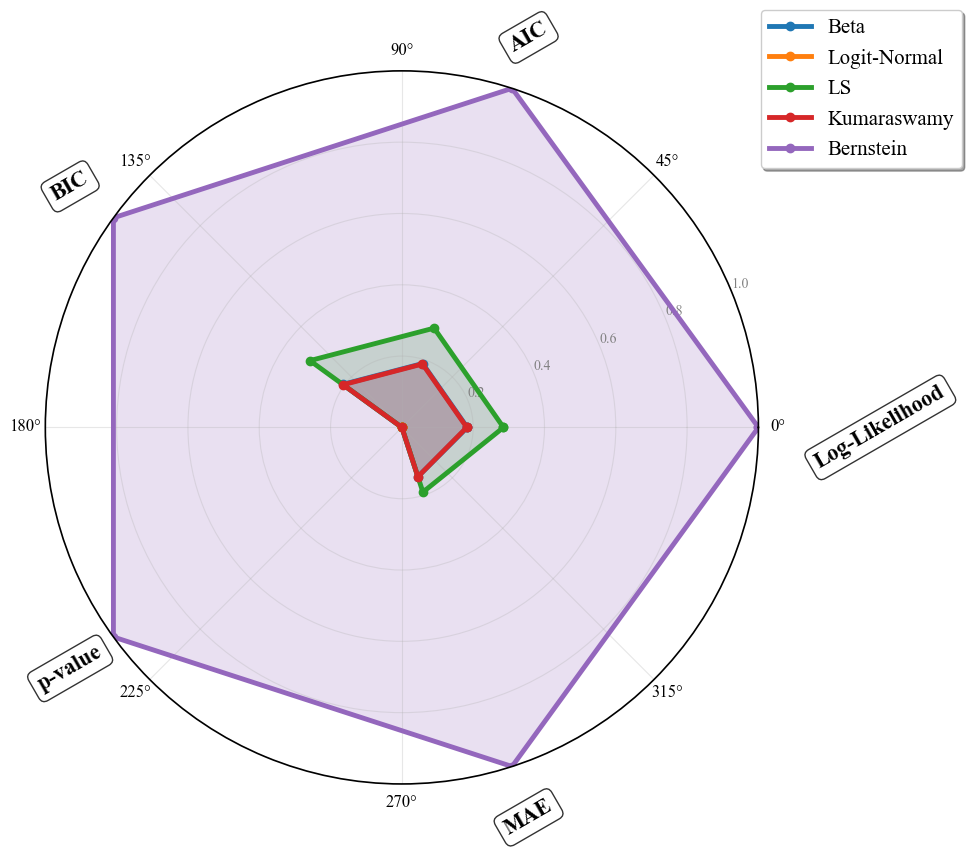}
    }
    \subfigure[$\;$ Bernstein approximation of Case ($\mbox{B}_1$)]
    {
        \centering
        \includegraphics[width=0.479\linewidth]{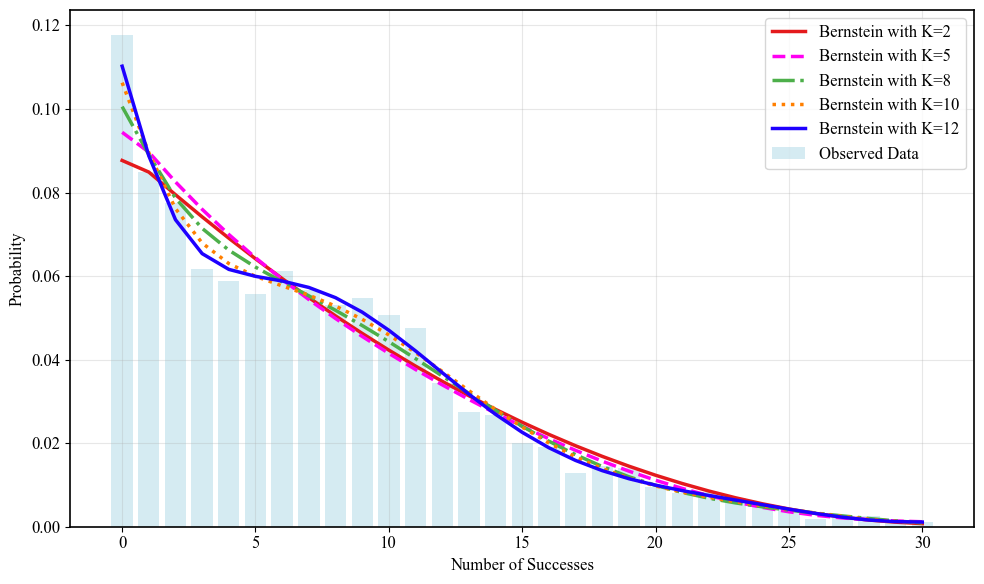}
    }
    \subfigure[$\;$ Bernstein approximation of Case ($\mbox{B}_3$)]
    {
        \centering
        \includegraphics[width=0.479\linewidth]{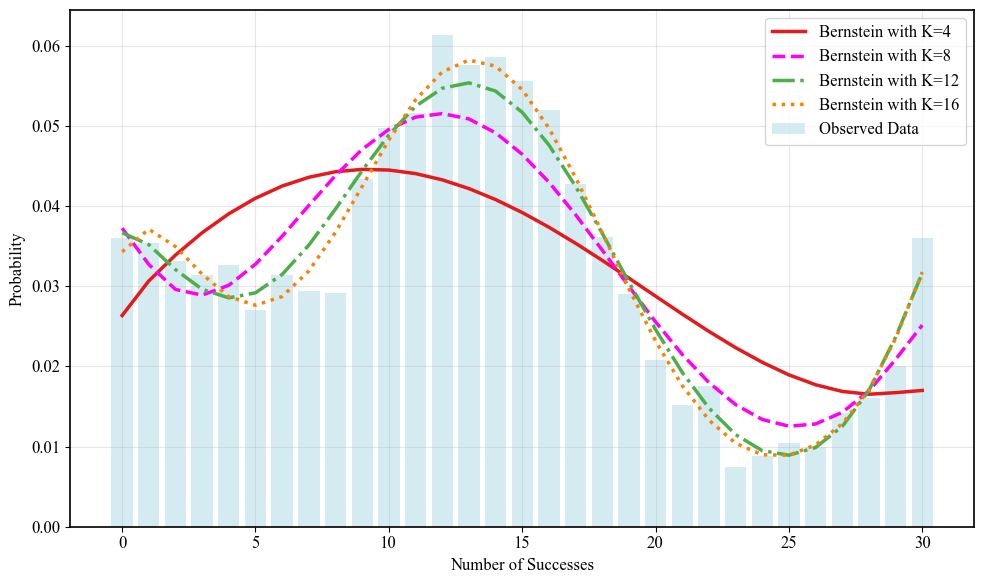}
    }
    \caption{$\;$ Model comparison for five distributions and Bernstein approximation plots.}
\end{figure}

\subsection{Bayesian inference with DA algorithm}       % Section 4.3

This section will systematically evaluate the performance of Bayesian inference based on DA algorithm in Bern-Bino distribution through numerical simulation. We will evaluate the performance of the posterior mean $\bar{\bla}$ as a point estimator, the coverage and accuracy of {\it Bayesian credible interval} (BCI), the convergence and sampling efficiency of DA algorithm and analyze the impact of different Dirichlet prior settings on the posterior inference.

We will construct multiple simulation scenarios to comprehensively evaluate the performance of the algorithm by controlling the sample size $n$ (Factor 1), the real distribution shape $\bla$ (Factor 2) and the Dirichlet prior parameter $\bal$ (Factor 3). For each scenario determined by Factor 1, 2 and 3, we will implement DA algorithm with the number of replications $R=$ 1,000 and collect evaluation indicators for statistical analysis.

Determine the Bernoulli trial number $m=15$ in binomial distribution, the following two different real parameter vectors will be considered:
\begin{namelist}{01234567}
\item[\hspace*{-0.05cm} Scenario 1 ($\mbox{S}_1$):] Symmetrical case with $\bla_\text{true} = (0.3, 0.4, 0.3)^{\T}$;

\item[\hspace*{-0.05cm} Scenario 2 ($\mbox{S}_2$):] Skewness with $\bla_\text{true} = (0.25, 0.45, 0.3)^{\T}$.
\end{namelist}
For each scenario, we generate datasets with various sample sizes of $n=200,500,800$ and perform Bayesian inference by using three different Dirichlet priors:
\begin{namelist}{01234567}
\item[\hspace*{-0.1cm} Prior 1 ($\mbox{P}_1$):] Non-informative prior with $\bal = \1$;

\item[\hspace*{-0.1cm} Prior 2 ($\mbox{P}_2$):] Correct prior with $\bal = 5 \times \bla_\text{true}$;

\item[\hspace*{-0.1cm} Prior 3 ($\mbox{P}_3$):] Error prior with $\bal = (5, 1, 5)^{\T}$.
\end{namelist}

For each experimental setup determined by $\{\mbox{S}, n, \mbox{P}, r\}$ in $r \in \bbN_{1:R}$, we run $M = 4$ independent Markov chains. Starting from different random initial values, each chain performs $G =$ 10,000 iterations, discards the previous $B = $ 2,500 as the burn-in period, and retains the last 7,500 samples for analysis. After the iteration, we combine the four chains to get a posterior sample with a size of 30,000. Based on the combined posterior samples, the posterior mean $\bar{\bla}$, bias, 95\% BCI, interval width, {\it effective sample size} (ESS) and Gelman--Rubin convergent diagnostic statistics ($\hR$) can be calculated. By independently repeating $R$ experiments, we can calculate the average of each index, the MSE of the posterior mean and the coverage rate \textsf{CR} of BCI.

\begin{figure}[H]
    \centering
    \subfigure[$\;$ Absolute bias and MSE]
    {
        \centering
        \includegraphics[width=1\linewidth]{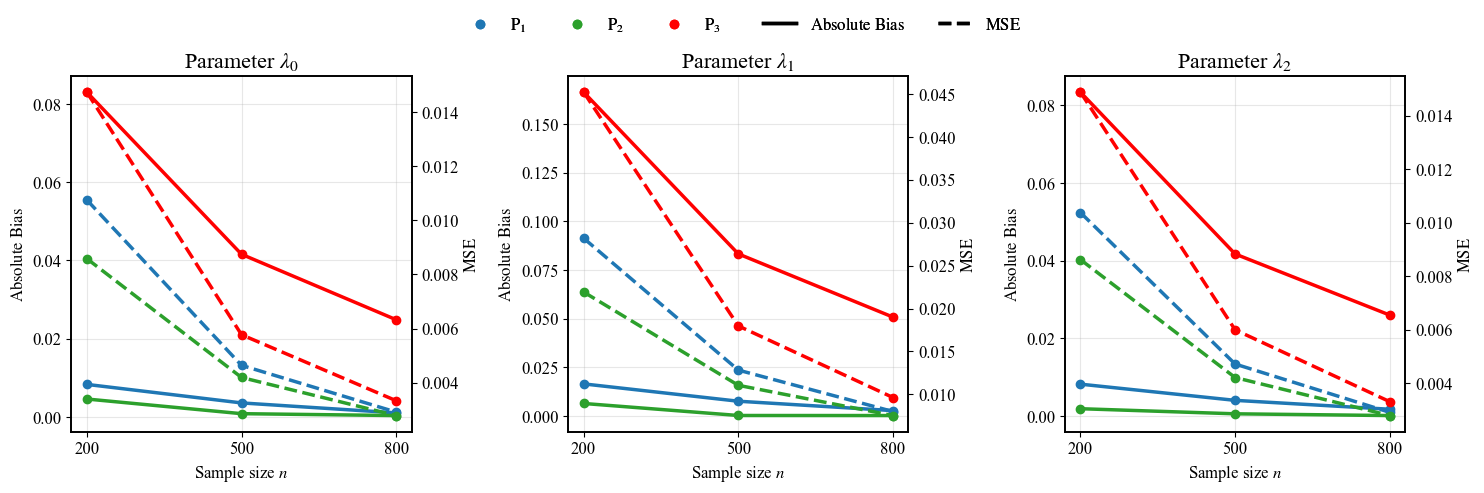}
    }
    \subfigure[$\;$ Coverage rate and interval width]
    {
        \centering
        \includegraphics[width=1\linewidth]{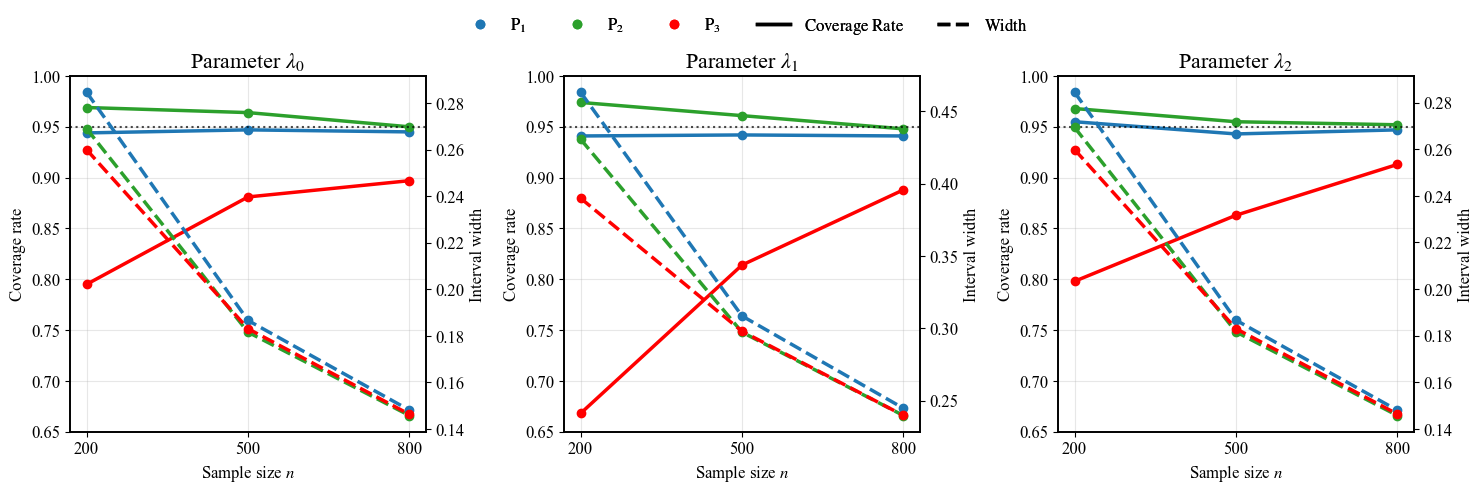}
    }
    \caption{$\;$ Performance indicators versus sample size under different priors for $\mbox{S}_1$.}
\end{figure}

Figure 4 shows us the absolute bias and MSE of the posterior mean $\bar{\bla}$, the coverage rate and interval width of 95\% BCIs with the sample size $n$ under different priors in Scenario 1. From the figure, we can found that the absolute bias and MSE decrease with the increase of sample size, and the correct prior information can improve the estimation accuracy especially in the case of small samples. For BCI, the interval width will decrease with the increase of the sample size, and the coverage rate of the correct prior ($\mbox{P}_2$) is the closest to the ideal value, while the coverage rate of the wrong prior ($\mbox{P}_3$) is obviously low when the $n$ is small, and the impact will weaken with the increase of $n$. Figure 5 presents the average coverage rate of $\bla$ for each prior under various sample sizes by employing the radar chart, where we can intuitively observe the above conclusions.

\begin{figure}[H]
    \centering
    \includegraphics[width=0.65\linewidth]{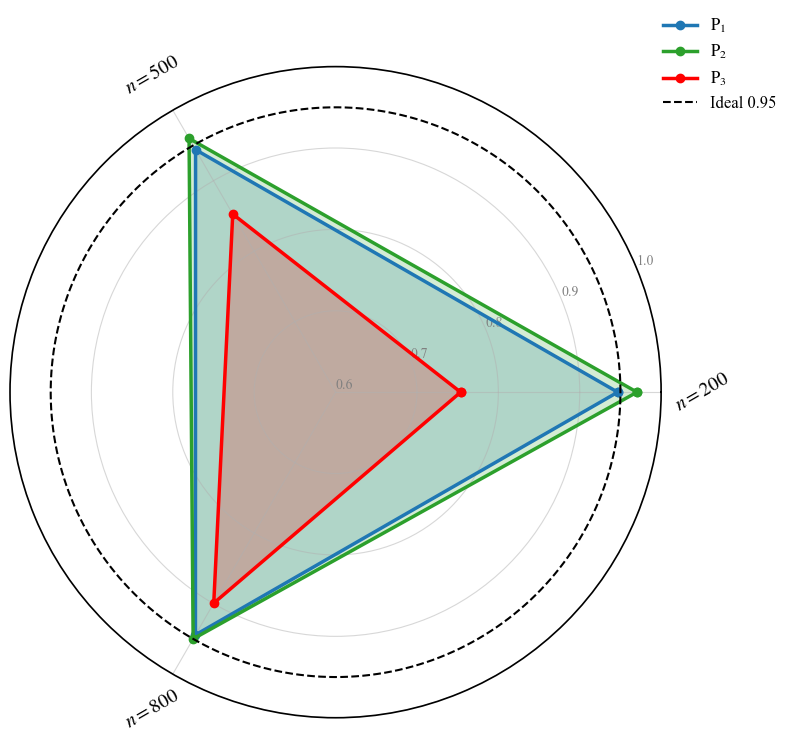}
    \caption{Radar chart of Coverage rate performance for $\mbox{S}_1$.}
\end{figure}

ESS is an indicator to measure the sampling efficiency of MCMC, which converts the MCMC sample chain with autocorrelation into an equivalent independent sample size, which is calculated based on the sum of the sample {\it autocorrelation function} (ACF):
$$
    \mbox{ESS} = \frac{N}{1 + 2 \sum_{t=1}^\infty \rho_t},
$$
where $N$ is the total number of MCMC samples and $\rho_t$ is the autocorrelation coefficient of samples with an interval of $t$. In practice, the sum is usually truncated when $\rho_t$ becomes insignificant. And the Gelman--Rubin statistic $\hat{R}$ (\cite{GR1992}) is used to diagnose the convergence of MCMC sampling. The statistic is constructed by comparing the inter chain variance ($B$) and intra chain variance ($W$) of multiple parallel Markov chains (\cite{BG1998}). We calculate the corrected total variance estimate
$$
    \widehat{\Var}(\bla) = \frac{N-1}{N} W + \frac{1}{N} B,
$$
then the Gelman--Rubin statistic is $\hat{R} = \sqrt{\widehat{\Var}(\bla)/W}$. When the $\hat{R}$ drops below 1.05 even 1.01, it is generally considered that convergence is sufficient.

\begin{figure}[H]
    \centering
    \includegraphics[width=1\linewidth]{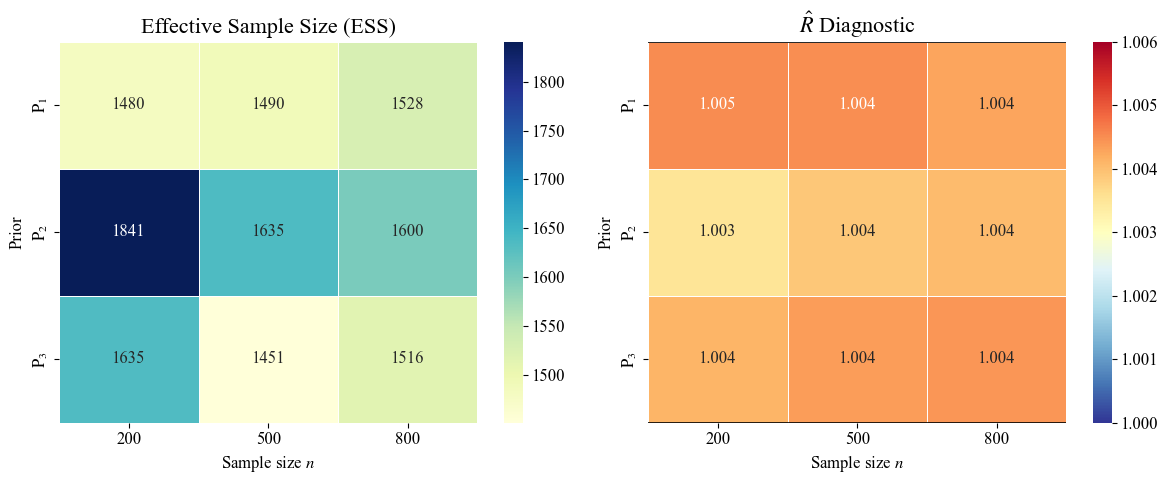}
    \caption{Heat map of DA algorithm efficiency for $\mbox{S}_1$.}
\end{figure}

Figure 6 exhibits the ESS and $\hat{R}$ values of DA algorithm under different sample sizes and priors by adopting heatmap. We can find that the ESS value is greater than $1,$$000$ and $\hat{R}$ is less than 1.01 for each case, which means that the efficiency and reliable convergence of posterior sampling and indicates DA algorithm will not be affected by initial value of sampling. In addition, ESS is significantly larger under the correct the prior and has higher sampling efficiency than other priors. For the scenario 2, we summarize the specific results of Bayesian inference in Table (\ref{tab3}), which demonstrates similar conclusions with scenario 1.

\newpage
\vkU
\baselineskip 0.10in
\begin{Table} \label{tab3} {\rm \hspace{-0.2cm}{\bf .}
Influence of different priors on Bayesian inference for $\mbox{S}_2$ based on $R = $ 1,000 replications under various sample sizes
\vspace{-0.15cm}
\begin{center}
\renewcommand{\arraystretch}{1.2} \tabcolsep 0.11in \doublerulesep 0.2pt
\begin{tabular}{|l|c|c|r|c|c|c|c|c|}  \hline\hline
    Sample size    & Parameter          & Prior        & \MC{1}{c|}{\textsf{Bias}} & \textsf{MSE}    & \textsf{CR} & \textsf{I-Wid} & \textsf{ESS}  & $\hat{R}$ \\ \hline
    \MR{9}{*}{200} & \MR{3}{*}{$\la_0$} & $\mbox{P}_1$ &    0.0059                 & 0.0099          & 0.954       & 0.277            & 1685          & 1.00      \\
    \MR{9}{*}{}    & \MR{3}{*}{}        & $\mbox{P}_2$ & $-$0.0014                 & 0.0084          & 0.958       & 0.261            & 2089          & 1.00      \\
    \MR{9}{*}{}    & \MR{3}{*}{}        & $\mbox{P}_3$ &    0.0945                 & 0.0166          & 0.739       & 0.258            & 1887          & 1.00      \\ \cline{2-9}
    \MR{9}{*}{}    & \MR{3}{*}{$\la_1$} & $\mbox{P}_1$ & $-$0.0136                 & 0.0279          & 0.959       & 0.465            & 1172          & 1.01      \\
    \MR{9}{*}{}    & \MR{3}{*}{}        & $\mbox{P}_2$ &    0.0050                 & 0.0229          & 0.957       & 0.428            & 1431          & 1.00      \\
    \MR{9}{*}{}    & \MR{3}{*}{}        & $\mbox{P}_3$ & $-$0.1890                 & 0.0538          & 0.589       & 0.404            & 1144          & 1.01      \\ \cline{2-9}
    \MR{9}{*}{}    & \MR{3}{*}{$\la_2$} & $\mbox{P}_1$ &    0.0076                 & 0.0108          & 0.955       & 0.290            & 1738          & 1.00      \\
    \MR{9}{*}{}    & \MR{3}{*}{}        & $\mbox{P}_2$ & $-$0.0036                 & 0.0094          & 0.953       & 0.271            & 2172          & 1.00      \\
    \MR{9}{*}{}    & \MR{3}{*}{}        & $\mbox{P}_3$ &    0.0946                 & 0.0174          & 0.761       & 0.269            & 1843          & 1.00      \\ \hline
    \MR{9}{*}{500} & \MR{3}{*}{$\la_0$} & $\mbox{P}_1$ &    0.0034                 & 0.0042          & 0.958       & 0.180            & 1733          & 1.00      \\
    \MR{9}{*}{}    & \MR{3}{*}{}        & $\mbox{P}_2$ & $-$0.0010                 & 0.0038          & 0.964       & 0.176            & 1885          & 1.00      \\
    \MR{9}{*}{}    & \MR{3}{*}{}        & $\mbox{P}_3$ &    0.0423                 & 0.0057          & 0.865       & 0.177            & 1788          & 1.00      \\ \cline{2-9}
    \MR{9}{*}{}    & \MR{3}{*}{$\la_1$} & $\mbox{P}_1$ & $-$0.0063                 & 0.0124          & 0.954       & 0.305            & 1208          & 1.00      \\
    \MR{9}{*}{}    & \MR{3}{*}{}        & $\mbox{P}_2$ &    0.0025                 & 0.0105          & 0.967       & 0.295            & 1306          & 1.00      \\
    \MR{9}{*}{}    & \MR{3}{*}{}        & $\mbox{P}_3$ & $-$0.0851                 & 0.0177          & 0.812       & 0.295            & 1200          & 1.01      \\ \cline{2-9}
    \MR{9}{*}{}    & \MR{3}{*}{$\la_2$} & $\mbox{P}_1$ &    0.0029                 & 0.0048          & 0.950       & 0.188            & 1783          & 1.00      \\
    \MR{9}{*}{}    & \MR{3}{*}{}        & $\mbox{P}_2$ & $-$0.0015                 & 0.0043          & 0.956       & 0.183            & 1946          & 1.00      \\
    \MR{9}{*}{}    & \MR{3}{*}{}        & $\mbox{P}_3$ &    0.0427                 & 0.0061          & 0.867       & 0.185            & 1796          & 1.00      \\ \hline
    \MR{9}{*}{800} & \MR{3}{*}{$\la_0$} & $\mbox{P}_1$ &    0.0017                 & 0.0027          & 0.944       & 0.143            & 1767          & 1.00      \\
    \MR{9}{*}{}    & \MR{3}{*}{}        & $\mbox{P}_2$ & $-$0.0016                 & 0.0026          & 0.952       & 0.141            & 1856          & 1.00      \\
    \MR{9}{*}{}    & \MR{3}{*}{}        & $\mbox{P}_3$ &    0.0303                 & 0.0034          & 0.872       & 0.142            & 1807          & 1.00      \\ \cline{2-9}
    \MR{9}{*}{}    & \MR{3}{*}{$\la_1$} & $\mbox{P}_1$ & $-$0.0028                 & 0.0077          & 0.947       & 0.242            & 1229          & 1.00      \\
    \MR{9}{*}{}    & \MR{3}{*}{}        & $\mbox{P}_2$ &    0.0025                 & 0.0073          & 0.939       & 0.237            & 1287          & 1.00      \\
    \MR{9}{*}{}    & \MR{3}{*}{}        & $\mbox{P}_3$ & $-$0.0586                 & 0.0103          & 0.857       & 0.237            & 1223          & 1.00      \\ \cline{2-9}
    \MR{9}{*}{}    & \MR{3}{*}{$\la_2$} & $\mbox{P}_1$ &    0.0011                 & 0.0029          & 0.946       & 0.149            & 1813          & 1.00      \\
    \MR{9}{*}{}    & \MR{3}{*}{}        & $\mbox{P}_2$ & $-$0.0009                 & 0.0028          & 0.941       & 0.147            & 1912          & 1.00      \\
    \MR{9}{*}{}    & \MR{3}{*}{}        & $\mbox{P}_3$ &    0.0283                 & 0.0035          & 0.900       & 0.147            & 1827          & 1.00      \\ \hline \hline
\end{tabular}
\end{center} }
\end{Table}
\baselineskip 0.30in 
%!TEX root = BSB.tex

\section{$\!\!\!\!\!\!\!$. Real data analysis}    % Section 5

\subsection{Women's mobility data}      % Section 5.1

A rural subsample of the Bangladesh Fertility Survey of 1989, conducted by the \textit{National Institute of Population Research and Training} (NIPORT), Dhaka (\cite{HC1990}). The data were subsequently compiled and analysed by \textcite{GMBS2002}. The data set is distributed with the \tsf{ltm} package for latent variable modelling (\cite{Rizopoulos2006}) and is loaded by \tsf{library(ltm)} and \tsf{data(Mobility)}.

This survey measures the social freedom of women in rural Bangladesh outside the household, containing 8445 samples. Each respondent was asked whether she was able to perform, alone, eight activities of daily life. The activities range from local movement to public participation, so their difficulty and their social acceptability differ substantially, and the degree of freedom varies considerably across women.

The raw data consist of eight binary item indicators. The variable used for model fitting is the total score
obtained by summing these indicators:

\vkU
\baselineskip 0.10in
\begin{Table} \label{zyf.BSB.tab4} {\rm  \ \  \hspace*{-0.5cm}{\bf .}
Data description of Bangladesh women's mobility dataset
\vspace{-0.3cm}
\begin{center}
\renewcommand{\arraystretch}{1.2} \tabcolsep 0.135in \doublerulesep 0.2pt
\begin{tabular}{llll}  \hline\hline
    \MC{1}{l}{Variable} & \MC{1}{c}{Value} & \MC{1}{c}{Description} & \MC{1}{c}{Role} \\ \hline
    Item 1 & 0/1 & Going to any part of the village, town or city & Item \\
    Item 2 & 0/1 & Going outside the village, town or city & Item \\
    Item 3 & 0/1 & Talking to an unknown man & Item \\
    Item 4 & 0/1 & Going to a cinema or cultural show & Item \\
    Item 5 & 0/1 & Going shopping & Item \\
    Item 6 & 0/1 & Going to a cooperative, mothers' club or other club & Item \\
    Item 7 & 0/1 & Attending a political meeting & Item \\
    Item 8 & 0/1 & Going to a health centre or hospital & Item \\
    score ($X$) & 0--8 & Number of activities that may be performed alone & Response \\ \hline\hline
\end{tabular}
\end{center} }
\end{Table}
\vspace{-0.3cm}
\noi  {\small   Note: For every item, 1 = the activity may be performed alone and 0 = it may not.}
\baselineskip 0.30in

\vkL
The score distribution is strongly over--dispersed, left--skewed and non--unimodal. The mode is attained at $X = 2$ (28.1 \%), while a clearly separated secondary mode appears at $X = 8$, i.e. 246 women (2.9 \%) may perform all eight activities alone. The distribution therefore mixes a large restricted subpopulation with a small fully unrestricted one, a structure that no unimodal prior on $p$ can represent.

The 8445 women are randomly split into a training set of $n_1 = 5912$ and a test set of $n_2 = 2533$. We perform the distributional fitting of binomial, Beta-Bino, LogitN-Bino, KM-Bino together with Bern-Bino distribution on training set and evaluate their fitting performance on test set.

\begin{figure}[H]
    \centering
    \includegraphics[width=0.65\linewidth]{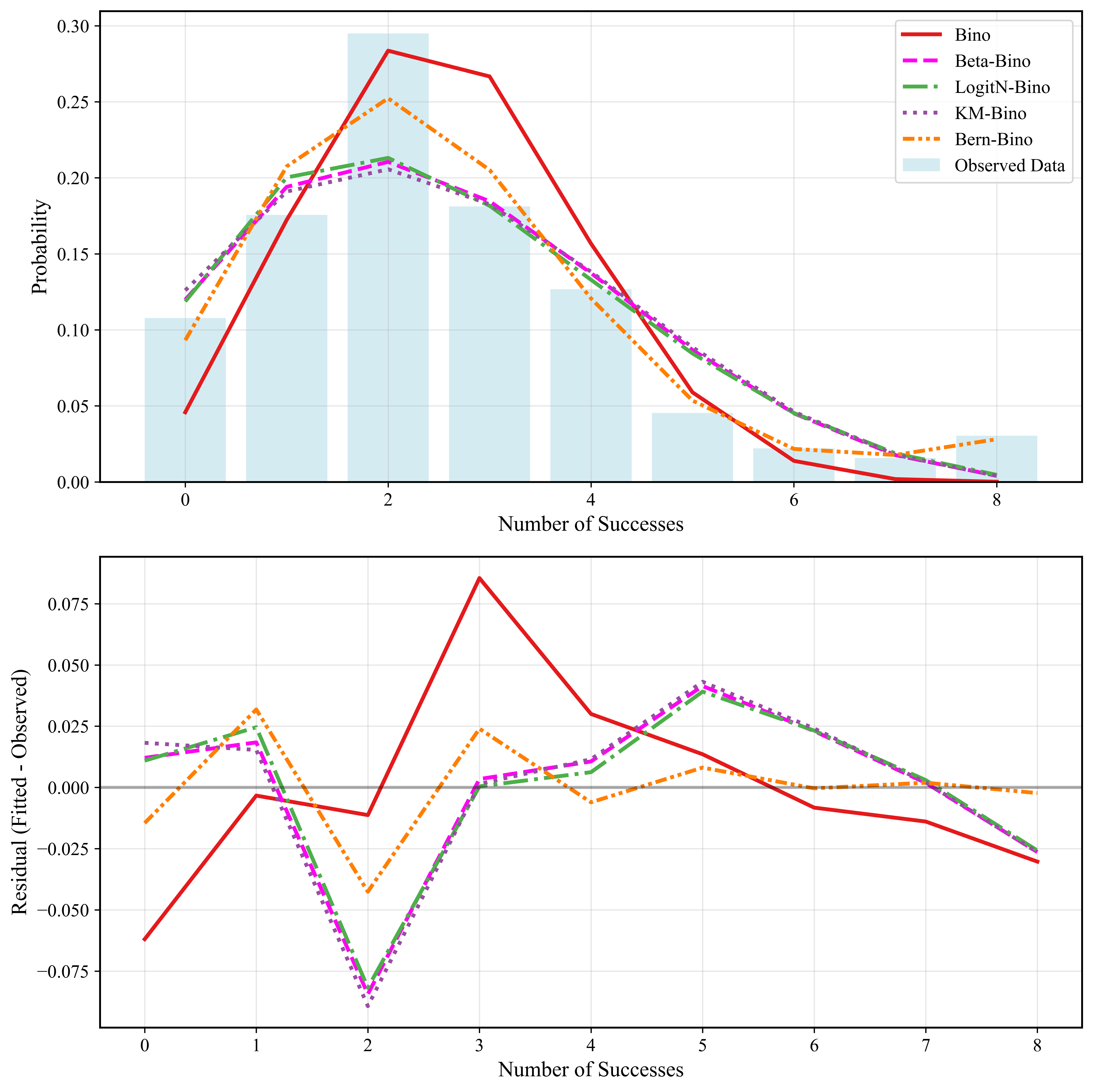}
    \caption{Comparison of the fitting performance for various distributions on the test set.}
    \label{zyf.BSB.fig7}
\end{figure}

Table \ref{zyf.BSB.tab5} reports the final test evaluation: the Bern-Bino distribution achieves the largest test log–likelihood ($-$4769.26, about 142 units higher than others) and the smallest test MSE and $L_1$ (about 0.36 and 0.61 times those of the best parametric distributions), while Figure \ref{zyf.BSB.fig7} shows that its fitted score distribution tracks the test frequencies much more closely than other distributions, especially at the secondary mode $X = 8$. Hence the advantage of the proposed model is genuine and holds on data that never participated in fitting.

\vkU
\baselineskip 0.10in
\begin{Table} \label{zyf.BSB.tab5} {\rm  \ \  \hspace*{-0.5cm}{\bf .}
Fitting performance for Bangladesh women's mobility score of five distributions on test set, respectively
\vspace{-0.2cm}
\begin{center}
\renewcommand{\arraystretch}{1.2} \tabcolsep 0.18in \doublerulesep 0.2pt
\begin{tabular}{llllll}  \hline\hline
    \MC{1}{c}{Distribution} & \MC{1}{c}{\tsf{Log-likelihood}} & \MC{1}{c}{\tsf{AIC}} & \MC{1}{c}{\tsf{BIC}} & \MC{1}{c}{\tsf{MSE}} & \MC{1}{c}{\tsf{TV}$^\dag$} \\ \hline
    Binomial                & $-$5285.79                      & 10573.57             & 10579.41             & 0.001504             & 0.1289                 \\
    Beta-Bino              & $-$4926.18                      & 9856.36              & 9868.03              & 0.001186             & 0.1108                 \\
    LogitN-Bino            & \bf{$-$4912.77}                 & \bf{9829.54}         & \bf{9841.21}         & \bf{0.001133}        & \bf{0.1075}            \\
    KM-Bino                & $-$4937.87                      & 9879.73              & 9891.41              & 0.001313             & 0.1158                 \\
    Bern-Bino              & \redbf{$-$4769.26}              & \redbf{9590.52}      & \redbf{9742.29}      & \redbf{0.000413}     & \redbf{0.0655}         \\ \hline\hline
\end{tabular}
\end{center} }
\end{Table}
\vspace{-0.3cm}
\noi  {\small   \tsf{TV}$^\dag$ means the total variation distance, which can be computed as $\sum_{i=1}^n |f(x_i) - \hat{f}(x_i)| / 2$.}
\baselineskip 0.30in

\vkL
The estimated Bernstein prior of the mobility level,
\begin{eqnarray*}
    \pi(\tau) &\ap& 0.01 \times \mBeta(\tau | 1, 26) + 0.16 \times \mBeta(\tau | 7, 20) + 0.78 \times \mBeta(\tau | 8, 19) \\ [2mm]
    && +\; 0.03 \times \mBeta(\tau | 24, 3)  + 0.02 \times \mBeta(\tau | 26, 1),
\end{eqnarray*}
concentrates about 94\% of the mass on a restricted subpopulation (mode around 0.26) and the remaining mass on a fully unrestricted subpopulation with $\tau$ close to 1, which directly reflects the secondary mode and is impossible for the beta, logit–normal and Kumaraswamy priors. In summary, on the Women’s Mobility data, the Bern-Bino distribution is best on the test set and provides a flexible, interpretable estimate of the prior density of $\tau$.

\subsection{MathExam14W data}       % Section 5.2

Individual results of the written final examination and of the accompanying online exercises of the course ``Mathematics 101" for first-year business and economics students at the University of Innsbruck, Austria. The data were collected by the Department of Statistics of the University of Innsbruck and are documented by \textcite{ZUL2014}, which is distributed with the \tsf{psychotools} package and is loaded by
\tsf{library(psychotools)} and \tsf{data("MathExam14W", package = "psychotools")}.

The written final examination consisted of $m = 13$ single-choice items and $n = 729$ students took the exam. The distribution of the number of solved items is unimodal and slightly right--skewed, with the mode at $X = 7$ and $X = 8$ (104 and 98 students, respectively). Both tails are long: 9 students solved no item at all and 32 students solved all 13 items. The item difficulties differ considerably, the proportion of correct answers ranging from 17.4 \% to 78.5 \%, which indicates substantial heterogeneity in ability and makes a simple binomial model inadequate. The regression analysis uses the response \tsf{nsolved} together with four covariates, namely \tsf{tests}, \tsf{gender}, \tsf{study} and \tsf{semester} as follows:

\vkU
\baselineskip 0.10in
\begin{Table} \label{zyf.BSB.tab6} {\rm  \ \  \hspace*{-0.5cm}{\bf .}
Data description of MathExam14W dataset
\vspace{-0.15cm}
\begin{center}
\renewcommand{\arraystretch}{1.2} \tabcolsep 0.04in \doublerulesep 0.2pt
\begin{tabular}{llll}  \hline\hline
    \MC{1}{l}{Variable} & \MC{1}{l}{Value} & \MC{1}{c}{Description} & \MC{1}{c}{Role} \\ \hline
    \tsf{nsolved} & 0--13 & Number of items solved correctly (the response X) & Response \\ \hdashline
    \tsf{tests} & 9--26 & \makecell[l]{Number of online test exercises solved correctly \\ before the examination, \\ a measure of study effort} & Covariate \\ \hdashline
    \tsf{gender} & female/male & Gender (326 female and 403 male students) & Covariate \\ \hdashline
    \tsf{study} & 155/571 & \makecell[l]{Type of degree programme: \\ 155 = four-year diploma programme (146 students) and \\ 571 = three-year bachelor programme (583 students)} & Covariate \\ \hdashline
    \tsf{semester} & 1--21 & Number of semesters enrolled in the programme & Covariate \\ \hline\hline
\end{tabular}
\end{center} }
\end{Table}
\baselineskip 0.30in

\vkL
To comprehensively evaluate the predictive performance and robustness of our proposed model, we randomly split the $n=729$ students into a training set and a test set with a 7:3 ratio. Consequently, the training set consists of 511 samples, while the remaining 218 samples are reserved for the test set.

\begin{figure}[H]
    \centering
    \includegraphics[width=0.5\linewidth]{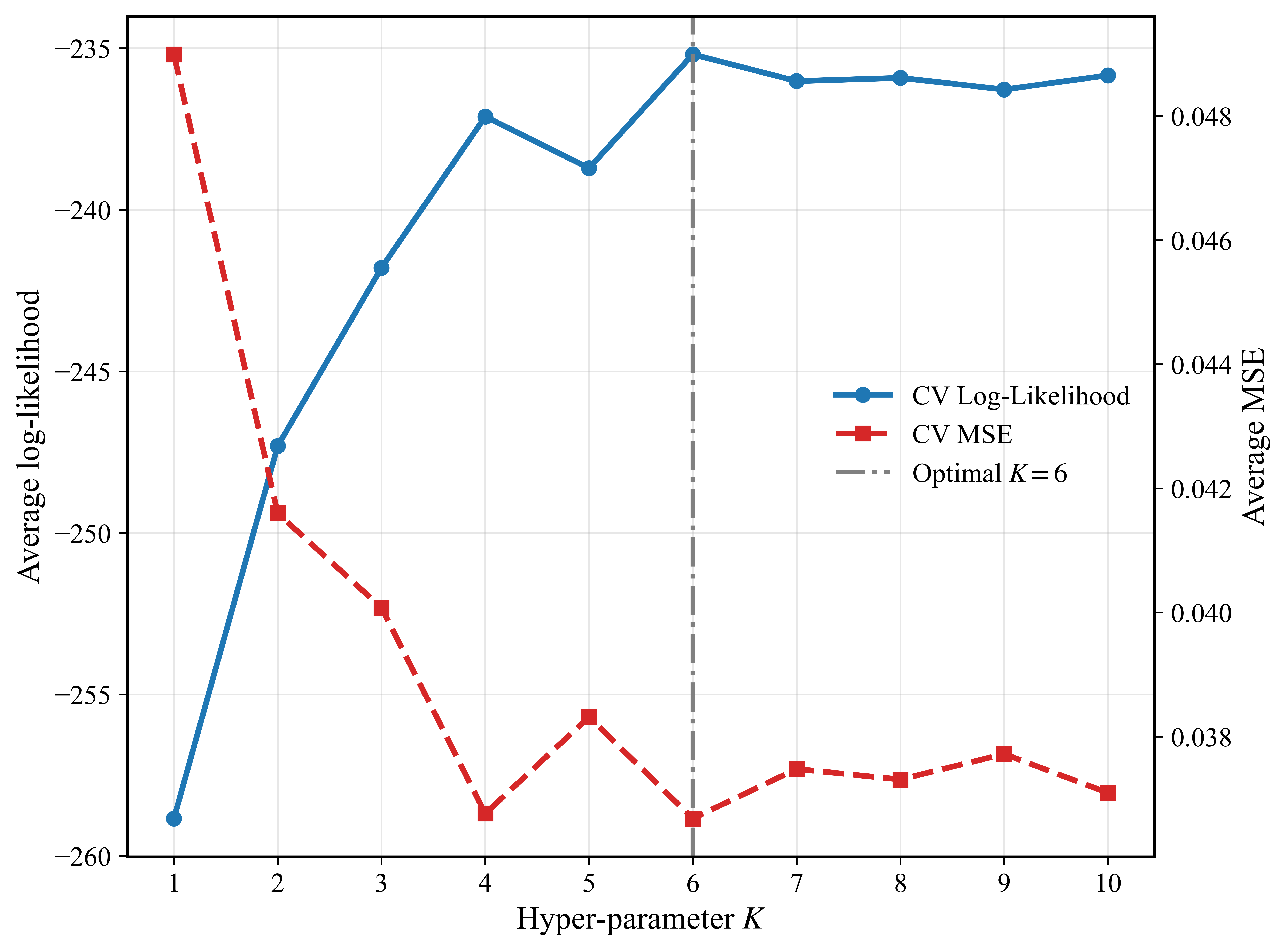}
    \caption{Cross--Validation for hyper-parameter $K$ selection.}
\end{figure}

We fit logistic regression, Beta-Bino regression (\cite{Prentice1986}), and our proposed Bern-Bino regression (with hyper-parameter $K = 6$ selected via cross--validation) on the training set. All three models describe binomial counts with a fixed number of trials, $X_i | p_i \sim \Binomial(13, p_i)$, and differ in the way the student--specific probability $p_i$ is specified:
\begin{namelist}{0123}
\item[\hspace*{-0.05cm} (i)] Logistic regression: $\logit(p_i) = \ibziT \bbe$: the success probability is common to all items of a student and varies across students only through the covariates;

\item[\hspace*{-0.05cm} (ii)] Beta-Bino regression: $p_i \sim \mBeta(a_i, b_i)$ with $\logit(\mu_i) = \ibziT\bbe$, where $\mu_i = a_i/(a_i + b_i)$, $a_i = \mu_i(1 - \rho)/\rho$ and $b_i = (1 -\mu_i)(1 - \rho)/\rho$ with a constant dispersion parameter $\rho$;

\item[\hspace*{-0.05cm} (iii)] Bern-Bino regression: $p_i \sim \Bernstein(K, \bla_i)$ with (\ref{zyf.BSB.2.12})--(\ref{zyf.BSB.2.13}) hold.
\end{namelist}

\vkU
\baselineskip 0.10in
\begin{Table} \label{zyf.BSB.tab7} {\rm  \ \  \hspace*{-0.5cm}{\bf .}
Comparisons of three regression models on test set
\vspace{-0.3cm}
\begin{center}
\renewcommand{\arraystretch}{1.2} \tabcolsep 0.12in \doublerulesep 0.2pt
\begin{tabular}{lllllll}  \hline\hline
    \MC{1}{l}{Model} & \MC{1}{c}{Number of Parameters} & \MC{1}{c}{Log-likelihood} & \MC{1}{c}{AIC}  & \MC{1}{c}{MSE}  & \MC{1}{c}{MAE} \\ \hline
    Logistic         & 5                              & $-$575.23                 & 1160.46         & \bf{0.04144}    & \bf{0.1668}    \\
    Beta-Bino       & 6                              & \bf{$-$516.40}            & \redbf{1044.81} & 0.04156         & 0.1669         \\
    Bern-Bino       & 16                             & \redbf{$-$507.32}         & \bf{1046.65}    & \redbf{0.03917} & \redbf{0.1621} \\ \hline\hline
\end{tabular}
\end{center} }
\end{Table}
\baselineskip 0.30in

\vkL
On the unseen test set, the Bern-Bino regression model unequivocally demonstrates superior predictive capability. It achieves the maximum out-of-sample log-likelihood alongside the MSE and MAE. Although the Beta-Bino model yields slightly better AIC and BIC scores---as information criteria heavily penalize the parameterized flexibility of the Bernstein prior---this metric is secondary in practical prediction tasks. The Bern-Bino regression model effectively trades a parsimonious structure for a powerful ability to capture unobserved heterogeneity, making it the most accurate and reliable approach for modeling complex binomial counts in real--world scenarios. 
%!TEX root = BSB.tex

\section{$\!\!\!\!\!\!\!$. Discussion}    % Section 6

In this paper, we introduced a Bernstein--prior--based generalization of the binomial mixture model, termed the Bern-Bino distribution, for bounded discrete counts that exhibit over--dispersion and nonstandard prior heterogeneity. The central idea is to replace the conventional beta prior with a finite beta mixture induced by Bernstein polynomials. Since Bernstein polynomials provide a uniform approximation to any continuous density on $[0, 1]$, the proposed model avoids the restrictive unimodality and prespecified shape of common parametric priors while retaining a tractable mixture structure.

From a methodological perspective, the proposed framework combines frequentist and Bayesian inference. The mixture representation yields closed--form moments and a simple stochastic representation; the MM algorithm provides a stable route to maximum likelihood estimation under the simplex constraint, while sandwich and bootstrap procedures produce interval estimates that remain useful in small samples. The covariate--dependent weights extend the model to a flexible regression setting, in which the latent prior can shift with subject characteristics rather than only through its mean. The simulation studies support both numerical reliability and modeling flexibility and real--data analyses provided external validation.

Several limitations still remain. The order $K$ is a model--selection parameter, the current cross--validation choice is practical but is not accompanied by automatic Bayesian treatment or asymptotic selection theory. The full--weight parameterization can be relatively expensive and is penalized by information criteria, and the Dirichlet prior on the weights is itself a subjective choice whose mis-specification can affect small--sample Bayesian intervals. Asymptotic confidence intervals near the simplex boundary also require further care. Future work could treat $K$ as unknown through reversible--jump MCMC or shrinkage priors, develop sparse and hierarchical extensions for multivariate or zero--inflated binomial data, establish posterior contraction and coverage theory for the Bern-Bino mixture, and investigate applications to high-dimensional genomics and social--survey data. Overall, the proposed model provides a principled, flexible, and computationally feasible framework for bounded count data whose latent success probability has a complex distribution.

%!TEX root = BSB.tex

%!TEX root = BSB.tex

\section*{Acknowledgement}      % Acknowledgement

Guo-Liang TIAN's research was partially supported by National Natural Science Foundation of China (No. 12171225). The work of Man-Lai TANG was partially supported by the Research Matching Grant (project: 700006 Applications of SAS Viya in Big Data Analytics), and the Big Data Intelligence Centre in The Hang Seng University of Hong Kong.

\renewcommand*{\bibfont}{\normalsize\baselineskip 0.30in}
% \nocite{*}
\printbibliography[title = {Reference}]

\appendix
%!TEX root = BSB.tex

\section{$\!\!\!\!\!\!\!$. Some Technical Derivations}      % Appendix A

\subsection{Asymptotic normality theorem for MLEs}       % Section A.1

Suppose that $\{X_i\}_{i=1}^n \iid F(x|\bth)$, the likelihood function is expressed as $\cL_n(\bth |\ibX)$ with $\ibX = (X_1, \ldots, X_n)^{\T}$. Then we calculate the MLE $\hbth_n$. To study the consistency and asymptotic normality of $\hbth_n$, we need a laundry list of assumptions, which has come to be known as the {\it Cram\'er--Rao} (CR) conditions.

\begin{theorem} \label{thm2} {\rm \textsf{(CR conditions)}.  Let $\{X_i\}_{i=1}^n$ and $X$ be i.i.d. r.v.s with pdf $f(x | \bth)$, which denotes the pmf for discrete case, and the corresponding \textit{cumulative distribution function} (cdf) be denoted by $F(x|\bth)$. In order to study the consistency and asymptotic normality of MLEs $\hbth_n$ of parameter vector $\bth$, we assume the following conditions:
\begin{namelist}{01234}
\item[\hspace*{-0.05cm} (A1)]   Identifiability, that is, $F(x|\bth_1) = F(x|\bth_2) \Longleftrightarrow \bth_1 = \bth_2$;

\item[\hspace*{-0.05cm} (A2)]   $\bth\in\bTh$ is an open interval in $\bbR^{d}$;

\item[\hspace*{-0.05cm} (A3)]   The observations are $\ibX = (X_1,\ldots,X_n)^{\T}$, where the $X_i$'s are i.i.d. either with pdf/pmf $f(x|\bth)$ for continuous case or discrete case;

\item[\hspace*{-0.05cm} (A4)]   The support $\mathbb{S} = \{x: \ f(x|\bth) > 0\}$ is free from $\bth$;

\item[\hspace*{-0.05cm} (A5)]   $\forall \ x \in \bbS$, $f(x|\bth)$ is differentiable with respect to $\bth$; that is, the likelihood function is smooth as a function of the parameter.
    \end{namelist}
Let $\bth_0 \in \bTh$ be the true value of $\bth$. Then there exists a sequence $\hbth_n$ such that $\hbth_n \rightarrow \bth_0$ in probability. Under the condition of consistency, Assumption (A6) and (A7) provide sufficient conditions for the asymptotic normality of $\hbth_n$.
\begin{namelist}{01234}
\item[\hspace*{-0.05cm} (A6)]   For all $x \in \bbS$ (specified in (A4)), the density $f(x|\bth)$ is three times differentiable with respect to $\bth$ and the third derivative is continuous. The corresponding integral (or sum in discrete case) of partial derivative
    $$
       \int \frac{\pa^3 f(x|\bth)}{\pa\theta_r \pa\theta_s \pa\theta_t} \rd x
    $$
    can be obtained by differentiating under the integral sign (or sum);

\item[\hspace*{-0.05cm} (A7)]   There exists a positive number $c(\bth_0)$ and a function $M_{\bth_0}(x)$ such that
    \begin{eqnarray*}
        \left|\frac{\pa^3}{\pa\theta_r \pa\theta_s \pa\theta_t}\log f(x|\bth)\right| \le M_{\bth_0}(x) \quad \mbox{for all} \quad x \in \bbS, \ ||\bth-\bth_0|| < c(\bth_0),
    \end{eqnarray*}
    and $E_{\bth_0}(M_{\bth_0}(X)) < \infty$.
\end{namelist}
Then any consistent sequence $\hbth_n$ satisfies
\begin{eqnarray*}
   \sqrt{n}\left(\hbth_n-\bth_0\right) \overset{\text{D}}{\rightarrow} N_d \left(\0, \bJ^{-1}(\bth_0)\right),
\end{eqnarray*}
where $\bJ(\bth)=[J_{ij}(\bth)]$ is the Fisher information matrix with
    \begin{equation*}
        J_{ij}(\bth) = -E\left[\frac{\pa^2}{\pa\theta_i\pa\theta_j} \log \cL_n(\bth|\ibX)\right]. \tag*{$\|$}
    \end{equation*}
}\end{theorem}

\subsection{Validation of CR conditions for Bern-Bino distribution}        % Section A.2

Suppose that $\{X_i\}_{i=1}^n \iid \BernBino(m, K, \bla)$ with $\bla \in \bbT_{K+1}^{\rm open}$, which satisfies Assumption (A3) in Theorem \ref{thm2} with discrete case. And the pmf of $X_i$ is given by
\begin{eqnarray*}
   \BernBino(x_i|m, K, \bla) \sr{(\ref{eqn2.2})} \binom{m}{x_i} \iba_i^{\T}\bla, \quad x_i \in \bbN_{0:m},
\end{eqnarray*}
where $\iba_i =(a_{i0}, a_{i1}, \dots, a_{iK})^{\T}$, the parameter $\bla = (\la_0, \dots, \la_K)^{\T}$ is identifiability (reason in Section 1.2) and $\bbT_{K+1}^{\rm open}$ is a open set in $\bbR^{K+1}$, which satisfies Assumption (A1), (A2). The support $\bbS_{X_i}= \bbN_{0:m}$ is free from $\bla$, therefore, the Assumption (A4) holds. The log-likelihood function of $\bla$ is
\begin{eqnarray*}
   \ell(\bla|\Y{obs}) \sr{(\ref{zyffour2.4})} \sum_{i=1}^n \log\binom{m}{x_i} + \sum_{i=1}^n \log \left( \iba_i^{\T}\bla \right) \qwi  \bla \in \bbT_{K+1}^{\rm open},
\end{eqnarray*}
which is a smooth function for $\bla_{-0}$ with partial derivatives
\begin{eqnarray*}
    \frac{\partial\ell(\bla|\Y{obs})}{\partial\la_r} &=& \sum_{i=1}^n \frac{a_{ir} - a_{i0}}{\iba_i^{\T}\bla} \qand \\ [2mm]
    \frac{\partial^3\ell(\bla|\Y{obs})}{\partial\la_r\partial\la_s\partial\la_t} &=& \sum_{i=1}^n \frac{2(a_{ir}-a_{i0})(a_{is}-a_{i0})(a_{it}-a_{i0})}{(\iba_i^{\T}\bla)^3},\quad r,s,t \in \bbN_{1:K}.
\end{eqnarray*}
Because pmf is a linear function of $\la_k$, which has good properties, so the log-likelihood function is third times differentiable and the derivative function is continuous. In the neighbor of true value $\bla_{-0,\rm true}$, we can choose a small enough constant $c(\bla_{-0,\rm true})$ such that all $\la_k \ge \delta > 0$. Therefore, we obtain the upper bound
\begin{eqnarray*}
    \bigg|\frac{\partial^3\ell(\bla|\Y{obs})}{\partial\la_r\partial\la_s\partial\la_t}\bigg| \le \frac{2c_{\max}^3}{(\delta c_{\min})^3}
\end{eqnarray*}
is a constant that does not depend on $x$ or the indicators $r,s,t$, where $c_{\max} = \max\{|a_{ik} - a_{i0}| \; \big| \; i \in \bbN_{1:n}, \ k \in \bbN_{0:K} \}$ and $c_{\min} = \min\limits_{1 \le i \le n, 0\le k\le K} a_{ik}$ are two constants. Then the three times partial derivative has upper bound function $M_{\bla_{-0,\rm true}}(x) = 2c_{\max}^3/(\delta c_{\min})^3$ with finite moment $E[M_{\bla_{-0,\rm true}}(X)] = 2c_{\max}^3/(\delta c_{\min})^3 < \infty$. Therefore, we verified Assumption (A5), (A6) and (A7) hold. In conclusion, for $\bla_{-0} \in \bbT_{K}^{\rm open}$, Bernstein--binomial distribution satisfy the Cram\'er–Rao conditions. 
%!TEX root = BSB.tex

\newpage
\section{$\!\!\!\!\!\!\!$. Some other results in article}      % Appendix B
\vkL
\baselineskip 0.10in
\begin{Table} {\rm \hspace{-0.2cm}{\bf .}
Influence of different priors on Bayesian inference for $\mbox{S}_1$ based on $R = $ 1,000 replications under various sample sizes
\vspace{-0.15cm}
\begin{center}
\renewcommand{\arraystretch}{1.2} \tabcolsep 0.11in \doublerulesep 0.2pt
\begin{tabular}{|l|c|c|r|c|c|c|c|c|}  \hline\hline
    Sample size    & Parameter          & Prior        & \MC{1}{c|}{\textsf{Bias}} & \textsf{MSE}    & \textsf{CR} & \textsf{I-Width} & \textsf{ESS}  & $\hat{R}$ \\ \hline
    \MR{9}{*}{200} & \MR{3}{*}{$\la_0$} & $\mbox{P}_1$ &    0.0083                 & 0.0108          & 0.944       & 0.285            & 1657          & 1.00      \\
    \MR{9}{*}{}    & \MR{3}{*}{}        & $\mbox{P}_2$ & $-$0.0046                 & 0.0086          & 0.969       & 0.269            & 2076          & 1.00      \\
    \MR{9}{*}{}    & \MR{3}{*}{}        & $\mbox{P}_3$ &    0.0830                 & 0.0147          & 0.795       & 0.260            & 1897          & 1.00      \\ \cline{2-9}
    \MR{9}{*}{}    & \MR{3}{*}{$\la_1$} & $\mbox{P}_1$ & $-$0.0165                 & 0.0282          & 0.941       & 0.463            & 1116          & 1.01      \\
    \MR{9}{*}{}    & \MR{3}{*}{}        & $\mbox{P}_2$ &    0.0065                 & 0.0219          & 0.974       & 0.431            & 1368          & 1.00      \\
    \MR{9}{*}{}    & \MR{3}{*}{}        & $\mbox{P}_3$ & $-$0.1660                 & 0.0452          & 0.668       & 0.390            & 1111          & 1.01      \\ \cline{2-9}
    \MR{9}{*}{}    & \MR{3}{*}{$\la_2$} & $\mbox{P}_1$ &    0.0082                 & 0.0104          & 0.955       & 0.284            & 1667          & 1.00      \\
    \MR{9}{*}{}    & \MR{3}{*}{}        & $\mbox{P}_2$ & $-$0.0019                 & 0.0086          & 0.968       & 0.269            & 2079          & 1.00      \\
    \MR{9}{*}{}    & \MR{3}{*}{}        & $\mbox{P}_3$ &    0.0834                 & 0.0149          & 0.798       & 0.260            & 1897          & 1.00      \\ \hline
    \MR{9}{*}{500} & \MR{3}{*}{$\la_0$} & $\mbox{P}_1$ &    0.0036                 & 0.0046          & 0.947       & 0.187            & 1668          & 1.00      \\
    \MR{9}{*}{}    & \MR{3}{*}{}        & $\mbox{P}_2$ &    0.0009                 & 0.0042          & 0.964       & 0.182            & 1839          & 1.00      \\
    \MR{9}{*}{}    & \MR{3}{*}{}        & $\mbox{P}_3$ &    0.0416                 & 0.0058          & 0.881       & 0.183            & 1632          & 1.00      \\ \cline{2-9}
    \MR{9}{*}{}    & \MR{3}{*}{$\la_1$} & $\mbox{P}_1$ & $-$0.0076                 & 0.0128          & 0.942       & 0.309            & 1134          & 1.01      \\
    \MR{9}{*}{}    & \MR{3}{*}{}        & $\mbox{P}_2$ & $-$0.0003                 & 0.0111          & 0.961       & 0.297            & 1234          & 1.00      \\
    \MR{9}{*}{}    & \MR{3}{*}{}        & $\mbox{P}_3$ & $-$0.0833                 & 0.0180          & 0.814       & 0.298            & 1082          & 1.01      \\ \cline{2-9}
    \MR{9}{*}{}    & \MR{3}{*}{$\la_2$} & $\mbox{P}_1$ &    0.0040                 & 0.0047          & 0.943       & 0.187            & 1668          & 1.00      \\
    \MR{9}{*}{}    & \MR{3}{*}{}        & $\mbox{P}_2$ & $-$0.0006                 & 0.0042          & 0.955       & 0.182            & 1833          & 1.00      \\
    \MR{9}{*}{}    & \MR{3}{*}{}        & $\mbox{P}_3$ &    0.0417                 & 0.0060          & 0.863       & 0.183            & 1639          & 1.00      \\ \hline
    \MR{9}{*}{800} & \MR{3}{*}{$\la_0$} & $\mbox{P}_1$ &    0.0011                 & 0.0029          & 0.945       & 0.148            & 1709          & 1.00      \\
    \MR{9}{*}{}    & \MR{3}{*}{}        & $\mbox{P}_2$ &    0.0004                 & 0.0028          & 0.950       & 0.146            & 1795          & 1.00      \\
    \MR{9}{*}{}    & \MR{3}{*}{}        & $\mbox{P}_3$ &    0.0249                 & 0.0033          & 0.897       & 0.146            & 1702          & 1.00      \\ \cline{2-9}
    \MR{9}{*}{}    & \MR{3}{*}{$\la_1$} & $\mbox{P}_1$ & $-$0.0028                 & 0.0080          & 0.941       & 0.245            & 1161          & 1.01      \\
    \MR{9}{*}{}    & \MR{3}{*}{}        & $\mbox{P}_2$ & $-$0.0003                 & 0.0075          & 0.948       & 0.240            & 1212          & 1.01      \\
    \MR{9}{*}{}    & \MR{3}{*}{}        & $\mbox{P}_3$ & $-$0.0508                 & 0.0096          & 0.888       & 0.240            & 1139          & 1.01      \\ \cline{2-9}
    \MR{9}{*}{}    & \MR{3}{*}{$\la_2$} & $\mbox{P}_1$ &    0.0017                 & 0.0029          & 0.947       & 0.148            & 1715          & 1.00      \\
    \MR{9}{*}{}    & \MR{3}{*}{}        & $\mbox{P}_2$ & $-$0.0001                 & 0.0028          & 0.952       & 0.146            & 1794          & 1.00      \\
    \MR{9}{*}{}    & \MR{3}{*}{}        & $\mbox{P}_3$ &    0.0260                 & 0.0033          & 0.913       & 0.146            & 1707          & 1.00      \\ \hline \hline

\end{tabular}
\end{center} }
\end{Table}
\baselineskip 0.30in 

\end{document}